\documentclass[11pt]{article}
\pdfoutput=1

\usepackage{euscript}
\usepackage{amssymb}
\usepackage{amsfonts}
\usepackage{amsbsy}
\usepackage{epsfig}
\usepackage{amsthm}
\usepackage{amscd}
\usepackage{amstext}
\usepackage{verbatim}
\usepackage{amsmath}
\usepackage{cancel}
\usepackage{capt-of}
\usepackage{empheq}
\usepackage{subfigure}
\usepackage{xcolor}

\usepackage{cite}

\usepackage[pdftex,linktoc=all]{hyperref}
\hypersetup{ 
colorlinks=true, 
linkcolor=black, 
citecolor=red, 
}

\def\ben{\begin{equation}}
\def\een{\end{equation}}

\let\a=\alpha   \let\d=\delta

\let\pa=\partial
\def\be{\begin{equation}}
\def\ee{\end{equation}}
\def\beq{\begin{equation}}
\def\eeq{\end{equation}}
\def\ba{\begin{array}}
\def\ea{\end{array}}

\def\dalemb#1#2{{\vbox{\hrule height .#2pt
       \hbox{\vrule width.#2pt height#1pt \kern#1pt
               \vrule width.#2pt}
       \hrule height.#2pt}}}

\newcommand{\bea}{\begin{eqnarray}}
\newcommand{\eea}{\end{eqnarray}}

\newcommand{\Tr}{{\rm Tr} }

\makeatletter
\newcommand*\bigcdot{\mathpalette\bigcdot@{.5}}
\newcommand*\bigcdot@[2]{\mathbin{\vcenter{\hbox{\scalebox{#2}{$\m@th#1\bullet$}}}}}
\makeatother

\newcommand{\specialcell}[2][c]{%
  \begin{tabular}[#1]{@{}c@{}}#2\end{tabular}}

\renewcommand{\eqref}[1]{(\ref{#1})}

\def \d {\partial}

\numberwithin{equation}{section}

\begin{document}
\frenchspacing
\begin{center}

{ \Large {\bf
Theory of hydrodynamic transport in \\
fluctuating electronic charge density wave states
}}

\vspace{1cm}

Luca V. Delacr\'etaz$^{\flat}$, Blaise Gout\'eraux$^{\flat \sharp \natural}$, Sean A. Hartnoll$^{\flat}$ and Anna Karlsson$^{\flat}$

\vspace{1cm}

{\small
$^{\flat}${\it Department of Physics, Stanford University, \\
Stanford, CA 94305-4060, USA }}

\vspace{0.5cm}

{\small
$^{\sharp}${\it APC, Universit\'e Paris 7, CNRS, CEA, \\ Observatoire de Paris, Sorbonne Paris Cit\'e, F-75205, Paris Cedex 13, France }}

\vspace{0.5cm}

{\small
$^{\natural}${\it
Nordita, KTH Royal Institute of Technology and Stockholm University, \\
Roslagstullsbacken 23, SE-106 91 Stockholm, Sweden}}

\vspace{1.6cm}

\end{center}

\begin{abstract}

We describe the collective hydrodynamic motion of an incommensurate charge density wave state in a clean electronic system. Our description simultaneously incorporates the effects of both pinning due to weak disorder and also phase relaxation due to proliferating dislocations. We show that the interplay between these two phenomena has important consequences for charge and momentum transport. For instance, it can lead to metal-insulator transitions. We furthermore identify signatures of fluctuating density waves in frequency and spatially resolved conductivities. Phase disordering is well known to lead to a large viscosity. We derive a precise formula for the phase relaxation rate in terms of the viscosity in the dislocation cores. We thereby determine the viscosity of the superconducting state of BSCCO from the observed melting dynamics of Abrikosov lattices and show that the result is consistent with dissipation into Bogoliubov quasiparticles.

\end{abstract}
\thispagestyle{empty}
\pagebreak
\pagenumbering{arabic}

\tableofcontents

\newpage


\section{Introduction}

\subsection{Hydrodynamics for metals}

Recent theoretical and experimental work has characterized anomalous hydrodynamic transport regimes in metals. These can arise due to the total momentum being long-lived, leading to a velocity field in the long wavelength hydrodynamic description. There are two key consequences of a long-lived velocity field for transport. The first are viscous effects, developed theoretically from various angles in works including \cite{Gurzhi63, Gurzhi68, KS06, PhysRevLett.103.025301, KS11, PhysRevLett.113.235901, Forcella:2014gca, Levitov2016, 2016arXiv160707269G}. Signatures of viscous electron flow have been observed in recent experiments \cite{PhysRevB.51.13389, Bandurin1055,Moll1061}. The second are anomalous relations between dc transport observables that are controlled by the long momentum relaxation timescale, discussed in works such as \cite{Hartnoll:2007ih, PhysRevB.75.245104, Hartnoll:2008hs, PhysRevB.78.115406, Hartnoll:2012rj, Mahajan:2013cja, PhysRevB.93.075426} and recently observed in \cite{Crossno1058}.

A second source of long-lived hydrodynamic modes are spontaneously broken symmetries. Our work will be concerned with hydrodynamic phenomena due to spontaneously broken translational symmetries in metals. These are closely intertwined with the velocity field effects mentioned above. In a clean system the hydrodynamics of translationally ordered states can be found in textbooks \cite{chaikin2000principles}. The textbook hydrodynamics can be `broken' in two different ways. Firstly, it has been understood for a long time that the incommensurate charge density wave Goldstone mode (the `sliding mode' or `phason') is both gapped and broadened by broken translation invariance \cite{LEE1974703, PhysRevLett.35.1399, PhysRevB.17.535,PhysRevB.19.3970,RevModPhys.60.1129}. Secondly, it has also been understood for some time that mobile dislocations lead to phase fluctuations that can ultimately melt the ordered state \cite{PhysRevB.19.2457,PhysRevLett.42.65} and have important effects on transport including a large viscosity \cite{PhysRevB.19.2457, PhysRevB.22.2514}. Our work will show that the combination of both pinning and phase fluctuations leads to novel hydrodynamic transport phenomena. In order for the (gapped, broadened and phase-fluctuating) Goldstone mode to make interesting contributions to transport it will be necessary to work in a clean limit where the effects of disorder are weak.

Unlike most previous analyses of density wave hydrodynamics, our discussion will not assume Galilean invariance. Galilean invariance is important (as we shall recall) for features such as a vanishing dc conductivity in pinned Wigner crystals \cite{RevModPhys.60.1129} and the absence of `climb' motion of dislocations \cite{PhysRevB.22.2514,Beekman:2016szb}. However, various non-Fermi liquid regimes that are likely unstable to translational order do not have Galilean-invariant effective low energy descriptions, for example many metallic quantum critical systems \cite{PhysRevB.14.1165}.

\subsection{Summary of results and their application}

There are several well-studied instances of spontaneous translational order in conventional electronic fluids. These include charge density waves due to the Peierls instability of a nested Fermi surface, Wigner crystals arising from potential energy dominance in dilute electron gases and Abrikosov lattices in superconductors. Translational order
is also found to be ubiquitous in strongly correlated systems such as doped Mott insulators. Our hydrodynamic theory of transport 
is independent of the mechanism that leads to translational order. We describe the `universal' (in the sense of low energy dynamics) consequences of 
simultaneously incorporating weak momentum relaxation and weak phase relaxation of translational order. After developing the formalism, three diverse applications are considered in this paper.

{\it Viscosity.---}Dislocations in charge density waves are vortices in the order parameter. The proliferation of dislocations melts the charge density wave and restores the broken symmetry \cite{PhysRevB.19.2457,PhysRevLett.42.65}. In terms of transport, the flow of dislocations render finite the otherwise infinite viscosities (bulk and shear) of a charge density wave state \cite{PhysRevB.19.2457, PhysRevB.22.2514}. We obtain a precise formula for the large finite viscosities of phase-disordered charge density wave states in terms of an effective shear viscosity $\eta^\text{eff}_\text{n}$ of the normal state in the dislocation core. For the simplest case of phase-disordered triangular biaxial charge density wave order the shear viscosity
\be\label{eq:sum1}
\eta = \frac{2}{x_v} \eta^\text{eff}_\text{n} \,.
\ee
Here $x_v < 1$ is the fraction of the area occupied by dislocation cores, see \S \ref{sec_memmat} below. Large viscosities should lead to significant hydrodynamic effects in electron flow of the kind measured in \cite{PhysRevB.51.13389, Bandurin1055, Moll1061}. In \S \ref{sec:melting} below we use \eqref{eq:sum1} to extract and understand the normal state viscosity from transport measurements of a phase-disordered Abrikosov lattice in optimally doped Bi$_2$Sr$_2$CaCu$_2$O$_8$ (BSCCO) \cite{1742-6596-150-5-052288}.

{\it Metal-insulator transitions and resistivity upturns.---}Another of our results will be a formula for the conductivity that incorporates the interplay between phase and momentum relaxation. This interplay is nontrivial: phase relaxation restores the spontaneously broken translational symmetry but momentum relaxation implies that this symmetry was never unbroken to begin with. For triangular biaxial charge density wave order, and further working with a nearly Galilean-invariant system to simplify the expression, we obtain the dc conductivity
\be\label{eq:sum2}
\sigma = \frac{n^2}{\chi_{\pi\pi}} \frac{\Omega}{\Omega \Gamma + \omega_o^2} \,.
\ee
See \S \ref{sec:cond} below. Here $\Omega$ is the phase relaxation rate, $\Gamma$ is the momentum relaxation rate and $\omega_o$ is the pinning frequency. The overall $n^2/\chi_{\pi\pi}$ is the usual Drude weight. In the formula \eqref{eq:sum2} both relaxation rates $\Omega$ and $\Gamma$ tend to increase with temperature. We show in \S \ref{sec:MIT} that this naturally leads to metal-insulator transitions by varying parameters that move the system closer to or further away from a charge density wave or crystalline state. We argue that \eqref{eq:sum2} may give a framework for understanding density-tuned metal-insulator transitions widely observed in large $r_s$ two-dimensional devices \cite{RevModPhys.82.1743} and the resistivity upturns widely seen at low temperatures in slightly underdoped cuprates, e.g. \cite{PhysRevLett.77.5417, PhysRevLett.81.4720, PhysRevLett.85.638, PhysRevLett.85.1738, 0295-5075-81-3-37008, Daou2009}.

 {\it Wavevector-dependent conductivity.---}The previous two paragraphs have discussed dc transport (i.e. with $\omega = k = 0$). Hydrodynamics furthermore fixes the inhomogeneous and time-dependent response at long wavelengths and low frequencies. In \S \ref{sec:sigmak} below we give the full frequency and wavevector dependence of the conductivity. In a certain limit this can be written as
\begin{equation}
\sigma(\omega,k)
	= \frac{n^2}{\chi_{\pi\pi}} \frac{\omega(\Omega-i\omega)}{\omega \left(\vphantom{\frac23}(\Gamma - i\omega)(\Omega-i\omega) + \omega_o^2\right) + \omega c^2 k^2 + i\Omega c_0^2 k^2} \, .
\end{equation}
Here $c$ and $c_0$ are certain sound speeds.
We will highlight various essential features of the $k$ dependence of this expression. These may be accessible to future experiments, giving a way to diagnose the presence of phase-disordered, fluctuating charge density wave dynamics.


\section{Density wave hydrodynamics with momentum and phase relaxation}\label{sec_relax}

In a phase with incommensurate translational order, spontaneous breaking of continuous translation symmetries leads to Goldstone modes in the low energy hydrodynamics \cite{chaikin2000principles}. In two space dimensions there are several possible symmetry breaking patterns. Either one or both generators of translations can be spontaneously broken. In a smectic, striped or charge density wave (CDW) phase, translations in a direction $\hat n$ are broken down to discrete translations, and a Goldstone phase $\phi_n$ parametrizes the quotient space $\mathbb R/\mathbb Z\simeq U(1)$. In the Wigner crystal (WC) phase, all translations are broken down to discrete translations generating a lattice $\Lambda$, and the quotient space $\mathbb R^2/\Lambda \simeq U(1)^2$ is parametrized by two Goldstone phases $\phi_i$. In the WC case there are different possible residual discrete symmetries of the lattice $\Lambda$, which constrain the form of the hydrodynamic constitutive relations. We focus on triangular WC (with 6-fold residual symmetry) and CDW phases.

The equations of hydrodynamics have four inputs that we will discuss in turn. Firstly, the free energy for the hydrodynamic degrees of freedom. Secondly, the conservation laws for conserved or approximately conserved densities. Thirdly, the `Josephson relations' for the time derivative of Goldstone modes. Finally, the constitutive relations for current densities.

\subsection{The free energy}

The fundamental hydrodynamic degrees of freedom are fluctuations $\varphi^A$ about thermal equilibrium. These are given by derivatives of the free energy $f$ with respect to their conjugate sources $s_A$. For density wave hydrodynamics these can be taken to be
\begin{align}
&{\rm WC}: && \varphi^A = \frac{\d f}{\d s_A} = \{\delta n,\, \delta s,\, \pi_\parallel,\, \pi_\bot,\, \lambda_\parallel,\, \lambda_\bot\}\, , && s_A = \{\delta \mu_e,\, \delta T,\, v_\parallel,\, v_\bot,\, s_\parallel,\, s_\bot \}\, ,\\
&{\rm CDW}: && \varphi^A = \frac{\d f}{\d s_A} = \{\delta n,\, \delta s,\, \pi^x,\, \pi^y,\, \d_n \phi_n\}\, , && s_A = \{\delta \mu_e,\, \delta T,\, v_x,\, v_y,\, s_n \}\, ,
\end{align}
where $n,\, s,\, \pi^i$ are the charge, entropy and momentum densities, with sources given by the chemical potential, temperature and velocity $\delta \mu_e,\, \delta T,\, v_i$. For the WC it is convenient to separate momentum density into its longitudinal and transverse components satisfying $\nabla \times \pi_\parallel = \nabla\cdot \pi_\perp = 0$. Also for the WC case the longitudinal and transverse parts of the Goldstone modes $\phi_i$ have been parametrized by $\lambda_\parallel = \nabla \cdot \phi$ and $\lambda_\perp = \nabla \times \phi$. Note that $s_\perp$, $s_\parallel$ and $s_n$ are sources, and should not be confused with the entropy $s$.

In appendix \ref{sec:hydroreview} we review the well-known description of the free energy for translational Goldstone bosons. Weak explicit breaking of translations gaps these fields, which become pseudo-Goldstone bosons. The free energy at wavevector $k$ is then
\begin{align}
&{\rm WC}: && 
f =   \frac{1}{2}\kappa |k\cdot \phi_k|^2 + \frac{1}{2}\mu (k^2+k_o^2) |\phi_k|^2 + \ldots \, ,  \qquad \qquad \label{f_relax} \\
&{\rm CDW}: && 
f = \frac{\kappa_n}{2} \left[ k_n^2 + k_o^2 \right] \phi_n^2 + \ldots \,. \qquad \qquad \label{f_relax_CDW}
\end{align}
Here $\kappa$ and $\mu$ are the bulk and shear moduli and $\kappa_n$ is the CDW modulus. $k_n = \hat n \cdot k$ and $k_o$ is the inverse correlation length of the phase at long distances: $\left<\phi(x)\phi(0)\right>\sim e^{-k_o\cdot x}$. The phase can be retained as a hydrodynamic variable so long as the correlation length is long compared to the electron mean free path. The weakly gapped Goldstone mode will be seen to have important consequences.

The free energies in the previous paragraph show how the sources $s_\perp$, $s_\parallel$ and $s_n$ are related to the linearized displacements (the Goldstone modes). At linear order, the remaining sources and conserved densities (charge, entropy, momentum) are likewise related by thermodynamic susceptibilities so that
\be\label{eq:sus}
\varphi^A = \chi^{AB} s_B \,.
\ee

\subsection{The (relaxed) conservation laws}

The charge and entropy conservation equations are:
\begin{equation}\label{conserv}
\dot n + \nabla\cdot j = 0\, , \qquad
\dot s + \nabla\cdot (j^Q/T) = 0\,.
\end{equation}
Entropy is conserved at the level of linear perturbations and is more convenient to work with than the energy, as it is directly conjugate to the temperature. Indeed $j^Q$ is the heat current.

Weak breaking of translation invariance means that momentum is not exactly conserved. The `conservation law' becomes
\begin{align}
&{\rm WC}: && 
\dot \pi^i + \nabla_j \tau^{ji} = -\Gamma\pi^i - \mu k_o^2 \phi^i + \ldots \, ,  \qquad \qquad\\
&{\rm CDW}: && 
\dot \pi^i + \nabla_j \tau^{ji} = -\Gamma^{ij}\pi_j - \kappa_n k_o^2n^i \phi_n + \ldots \,. \qquad \qquad
\end{align}
Here $\Gamma$ is the momentum relaxation rate, the $\ldots$ contains higher derivative corrections\footnote{Such as a term proportional to $\d^i\mu_e$. Such terms can be seen to have no other effect than redefining thermodynamic quantities such as $n$ or $s$.}, and the  $\phi^i$ term on the right hand side is imposed by Onsager relations (this is seen a posteriori once the full hydrodynamic equations are assembled). The only difference for the CDW relative to the WC case is that the reduced symmetry allows for a more general momentum relaxing term: $\Gamma^{ij} = \Gamma_n n^in^j + \Gamma_{\perp n} (\delta^{ij} - n^in^j)$. Recall that $\hat n$ is the direction of the CDW.

As with the pseudo-Goldstone bosons, the momenta can be retained as hydrodynamic variables so long as the momentum relaxation rate is small compared to the local thermalization rate.

\subsection{The (relaxed) Josephson relations}

In analogy to the case of superfluids, we will refer to equations for the time derivatives of $\phi_i$ (or $\phi_n$)
as `Josephson relations'. These equations follow directly from the fact that translations generate shifts in the Goldstone boson -- see the commutation relation \eqref{WC_com} -- together with the fact that the Hamiltonian can be written $\mathcal H= \pi^i v_i$ + \ldots, where the $\ldots$ only contain derivatives of the momentum. That is, in thermodynamic equilibrium $\dot \phi_i = v_i$. This mirrors the case of superfluids, where the phase is conjugate to the charge density $n$, so that $\mathcal H = \mu_e n +\ldots$ leads to the usual Josephson relation $\dot \phi = - \mu_e $ in equilibrium.

Phase relaxation rates $\Omega_\parallel$ and $\Omega_\perp$ can be introduced independently of momentum relaxation and appear in the Josephson relations. In order for the pseudo-Goldstone modes to survive as hydrodynamic modes, the phase relaxation rates must be small compared to the local equilibration rate. These Josephson relations become, for the WC
\begin{subequations}\label{WC_joseph_relax}
\begin{align}
(\d_t + \Omega_\parallel) \lambda_\parallel		\label{WC_joseph_relax_para}
	&= \nabla\cdot v + \ldots \, ,\\			\label{WC_joseph_relax_perp}
(\d_t + \Omega_\bot) \lambda_\perp
	&= \nabla \times v + \ldots \, ,
\end{align}
\end{subequations}
and for the CDW
\begin{equation}\label{CDW_phase_relax}
(\d_t + \Omega)\d_n \phi_n =\d_n v_n+\ldots \, .
\end{equation}
Here the $\ldots$ again refer to higher derivative corrections. The higher derivative corrections to the Josephson relations above will play no significant role in our main discussion, they are given in (\ref{joseph}) of Appendix \ref{app_S}. 
It is the gradients $\lambda_\parallel$ and $\lambda_\perp$ of the Goldstone modes that are physical and hence phase relaxation is introduced after taking a gradient of the Josephson relations. It is well-understood that phase relaxation arises due to the motion of dislocations, that act as vortices in the phases. We review this physics in Appendix \ref{app:dislocation}. In general the phase relaxation rates $\Omega_\parallel$ and $\Omega_\bot$ are distinct. In \S \ref{ssec_memmat_noP} we see that when translations are explicitly broken these phase relaxation rates acquire a mild $k$ dependence and become equal on the very longest wavelengths with $k \ll k_o$.

\subsection{The constitutive relations}

The constitutive relations express the currents of the conserved densities $j, j^Q$ and $\tau$ as a gradient expansion in terms of the conserved densities themselves (or their conjugate sources, which are linearly related to the densities according to (\ref{eq:sus}) and the discussion directly above, and in some places turn out to be more convenient). To first order in derivatives for the WC we can write
\begin{subequations}
\begin{align}
j &= n v - \sigma_0\nabla \mu_e - \alpha_0 \nabla T - \gamma_1\nabla s_\parallel + \ldots \, , \\
j^Q/T &= s v - \alpha_0\nabla \mu_e - (\bar\kappa_0/T) \nabla T - \gamma_2 \nabla s_\parallel + \ldots \, , \\
\tau^{ij} &=  \delta^{ij} \left(p-(\kappa+\mu)\lambda_\parallel\right) - \epsilon^{ij} \mu\lambda_\perp - \sigma^{ij} + \ldots \, .\label{tau_ij_consti}
\end{align}
\end{subequations}
The non-dissipative (or `reactive') 0-derivative terms in the above are fixed by the absence of entropy production and do not presuppose Galilean invariance or other such symmetries, see \cite{Hartnoll:2016apf} and appendix \ref{app_S}. The 1-derivative terms appear with transport coefficients $\{\sigma_0, \a_0,\bar\kappa_0, \gamma_1, \gamma_2 \}$. The 1-derivative corrections $\sigma^{ij}$ to the stress tensor are given by
\begin{equation}
\sigma^{ij} = \zeta \delta^{ij} (\nabla \cdot v) + \eta \left[\nabla^{i}v^{j} + \nabla^{j}v^{i} - \delta^{ij} (\nabla \cdot v)\right]\, ,
\end{equation}
where $\zeta$ and $\eta$ are the bulk and shear viscosities, respectively.

The above expressions are constrained by Onsager relations, which fix for example the 0-derivative contribution of $\lambda_\parallel$ and $\lambda_\perp$ to the stress tensor, and relate the $\gamma_i$ to coefficients that appear as higher derivative corrections to the Josephson relation (given in Appendix \ref{app_S}). The stress tensor $\tau^{ij}$ above is not symmetric, which it should be due to rotation invariance of the underlying symmetric state. However, it can be made symmetric by adding a transverse term which does not contribute to the conservation equation \eqref{conserv} and thus does not affect our computation of hydrodynamic modes and Green's functions below\footnote{
Specifically, 
\begin{equation*}
\tau'^{ij} = \delta^{ij}p + (\kappa+\mu)\delta^{ij}(\d\cdot \phi) + 2\mu \left[\d_{(i}\phi_{j)}-\delta_{ij}(\d\cdot \phi)\right] - \sigma^{ij}\, 
\end{equation*}
is symmetric and $\d_i(\tau'^{ij}-\tau^{ij})=0$.
}.

For a CDW the constitutive relations are
\begin{subequations}\label{consti}
\begin{align}
j &= n v - \sigma_0\nabla \mu_e - \alpha_0 \nabla T - \gamma_1 \hat n\d_n s_n + \ldots\, , \\
j^Q/T &= s v - \alpha_0\nabla \mu_e - (\bar\kappa_0/T) \nabla T - \gamma_2 \hat n\d_n s_n + \ldots\, , \\
\tau^{ij} &= \delta^{ij}p -n^i n^j \kappa_n \d_n \phi_n - \sigma^{ij} + \ldots\, ,
\end{align}
\end{subequations}
where the transport coefficients or `incoherent conductivities' are now matrices, e.g.
\begin{equation}
(\sigma_0)^{ij} = \sigma_0^n\, n^in^j + \sigma_0^{\perp n}\, (\delta^{ij} - n^in^j)\, .
\end{equation}
Because rotations are spontaneously broken, constitutive relations, which are really expectation values of operators, need not respect them. The 1-derivative corrections to the stress tensor now have the general form
\begin{equation}\label{CDW_viscosities}
\sigma^{ij} = \eta^{ijkl} \d_k v_l\, .
\end{equation}
A CDW in an isotropic system has residual 2-fold and reversal symmetries, which allow for 5 independent terms in $\eta^{ijkl}$ \cite{PhysRevA.6.2401}.

\subsubsection*{No Galilean invariance}

As mentioned in the introduction, throughout we purposely do not impose Galilean invariance, which would require $j \propto \pi$ as an operator relation and thus set $\sigma_0=\alpha_0= \gamma_1 = 0$. Unlike conventional fluids, electron fluids live in a system with a preferred frame, and a priori have no reason to be Galilean invariant. Galilean invariance can be strongly broken by interactions even in a translationally invariant low energy effective theory. Examples are close to quantum critical points in metals, where the fermions couple strongly to gapless bosons \cite{PhysRevB.14.1165}.

The Galilean limit will be discussed in various places throughout, including in the context of relaxation due to dislocations (\S \ref{sec_memmat}), where it qualitatively affects certain observables.

\subsection{Conductivities}
\label{sec:cond}

The free energy, conservation laws, Josephson relations and constitutive relations given above lead to a closed set of equations for the hydrodynamic variables $\varphi^A$. Solving these equations in the absence of external sources gives the hydrodynamics modes of the system. The modes themselves are not our primary interest, although of course they physically underly all our results for transport. We review the well-known modes of the clean system in Appendix \ref{ssec_modes}.

The conductivities that describe heat and charge transport are defined by Kubo formulae, in terms of the retarded Green's functions for the conserved densities. These Green's functions are computed from the hydrodynamic equations of motion using the method of Kadanoff and Martin \cite{KADANOFF1963419}. According to this method, one first puts the hydrodynamic equations into the form
\be\label{eq:heqns}
\dot \varphi^A + M^{AB}(k) s_B = 0\,.
\ee
The matrix $M$ is straightfowardly obtained by combining the free energy, conservation laws, Josephson relations and constitutive relations. With this matrix at hand, the matrix of retarded Green's functions for the $\varphi^A$ variables is given by
\be
G^R(\omega,k) = M(k) \frac{1}{i \omega \chi - M(k)} \chi \,.
\ee
Here the static susceptibilities $\varphi^A = \chi^{AB} s_B$. These Green's functions manifestly have poles on the hydrodynamic modes where the equations (\ref{eq:heqns}) are satisfied. 

\begin{figure}[h]
\centerline{
\includegraphics[width=0.6\linewidth,angle=0]{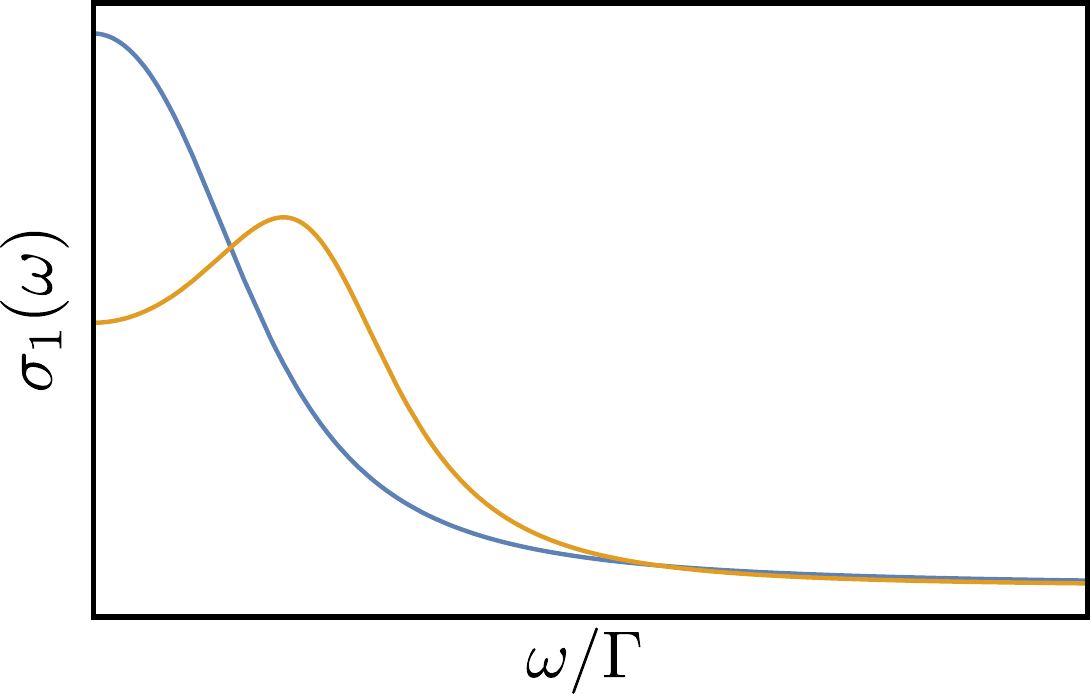}}
\caption{\label{fig_conductivities} Real part of the optical conductivity perpendicular (blue) and parallel (yellow) to the CDW. The yellow curve also illustrates the isotropic conductivity of a triangular WC. The peak occurs at nonzero frequency as long as $\omega_o^2 > \Omega^3/(\Gamma + 2\Omega)$. }
\end{figure}

The electrical conductivity is given by $\sigma(\omega) = \lim_{k\to 0} i \omega/k^2 \times G^R_{nn}(\omega,k)$. Using the Kadanoff-Martin Green's functions for a triangular WC, the conductivity is isotropic and given by\footnote{The Kadanoff-Martin procedure gives the longitudinal conductivity. It is also possible to compute the transverse conductivity (defined as a two-point function of transverse currents),
\begin{equation*}
\sigma_\perp(\omega,k\to 0)  =
\sigma_0 + \frac{n^2}{\chi_{\pi\pi}} \frac{\Omega_\perp - i\omega
}{(\Omega_\perp-i\omega)(\Gamma-i\omega)+ \omega_o^2}\, .
\end{equation*}
Locality requires this transverse response to equal the longitudinal response \eqref{WC_cond} when $k\to 0$, which implies $\Omega_\perp(k\ll k_o)=\Omega_\parallel(k\ll k_o)$. That is, we can write $\Omega_\bot = \Omega_\parallel \equiv \Omega$ when discussing $k=0$ conductivities. The $k$-dependence of the relaxation rates is further discussed in \S \ref{ssec_memmat_noP}.
}
\begin{equation}\label{WC_cond}
{\rm WC:}\qquad\sigma(\omega) = \sigma_0 + \frac{n^2}{\chi_{\pi\pi}} \frac{\Omega_\parallel - i\omega
}{(\Omega_\parallel-i\omega)(\Gamma-i\omega)+ \omega_o^2} \,,
\end{equation}
where we defined the `pinning frequency'
\be\label{eq:pin}
\omega_o^2 \equiv \frac{\mu k_o^2}{\chi_{\pi\pi}} \,.
\ee
In \eqref{WC_cond} we dropped frequency-independent terms that appear in the numerator but are subleading in the regime where $\omega, \omega_o, \Omega$ and $\Gamma$ are small.\footnote{In particular (\ref{eq:appgamma}) implies that a certain coefficient $\gamma_1 \to 0$ as $\Omega \to 0$, making a term proportional to $\gamma_1$ manifestly subleading.}
The pole structure in the conductivity is richer than a simpler Drude peak and arises because translations are spontaneously broken by the WC in addition to having weak explicit breaking. The frequency-dependent conductivity \eqref{WC_cond} is sketched in figure \ref{fig_conductivities}. A sufficiently large gap $\omega_o$ `pins' the collective mode, requiring a nonzero excitation energy, while $\Gamma$ and $\Omega$ determine the lifetime.

For a CDW, the conductivity along the CDW wavevector has the same form as the WC conductivity. In the direction perpendicular to the CDW, however, translations are not spontaneously broken and the effect of momentum relaxation is as in regular hydrodynamics, leading to a Drude-like conductivity:
\begin{subequations}\label{CDW_cond}
\begin{empheq}[left={\rm CDW}:\qquad\empheqlbrace\quad, right=\qquad]{align}
\sigma^n(\omega)&=\sigma_0^n+\frac{n^2}{\chi_{\pi\pi}} \frac{\Omega-i\omega 
}{ (\Omega-i\omega)(\Gamma_n-i\omega) + \omega_o^2}\, ,\\
\sigma^{\bot n}(\omega) &= \sigma_0^{\bot n} + \frac{n^2}{\chi_{\pi\pi}} \frac{1}{\Gamma_{\perp n} - i\omega}\, . \label{cond_y}
\end{empheq}
\end{subequations}
The CDW pinning frequency is defined as $\omega_o^2 \equiv \kappa_n k_o^2/\chi_{\pi\pi}$.
We again dropped a term in the numerator of $\sigma^n$, subleading when $\omega, \omega_o, \Omega$ and $\Gamma_n$ are small.

The pole structure of the conductivities without phase relaxation has been known for a long time (see e.g.~\cite{RevModPhys.60.1129}), but restricted to $\Omega =  \sigma_o = 0$. Our expressions are derived from a fully consistent hydrodynamics and incorporate the interplay of momentum and phase relaxation. Some phenomenological consequences
of these formulas for the conductivity are discussed in \cite{Delacretaz:2016ivq}. In particular, it is noted in \cite{Delacretaz:2016ivq} that peaks similar to those in figure \ref{fig_conductivities} are widely observed in bad metals. In \S\ref{sec:MIT} below we will show that the dc limit of these formulae give a simple mechanism for metal-insulator transitions.

The other thermoelectric conductivities have similar expressions to those above, with the same pole structure. For the WC, they can be expressed as
\begin{equation}\label{relaxed_cond}
\Sigma(\omega) = \Sigma_0 + \frac{(\Omega - i\omega)(A_\Sigma/\chi_{\pi\pi}) + \omega_o^2 B_\Sigma
}{(\Omega-i\omega)(\Gamma-i\omega)+\omega_o^2 } \,,
\end{equation}
with $\Sigma=\sigma,\,\alpha,\,(\bar\kappa/T)$ -- the electric, thermoelectric and thermal conductivities -- and
\begin{gather}\begin{aligned}
A_\sigma&= n^2\, , &
A_\alpha&= n s\, , &
A_{\bar\kappa/T}&=s^2\, ,\quad \\
B_\sigma&
	= 2n\gamma_1\, ,\quad &
B_\alpha&
	= n \gamma_2 + s\gamma_1\, , &
B_{\bar\kappa/T}&
	= 2s\gamma_2\, .
\end{aligned}\end{gather}
In (\ref{relaxed_cond}) we kept the (subleading) terms proportional to the coefficients $\gamma_1, \gamma_2$ for completeness.

It is important to note that in absence of any explicit translation symmetry breaking, so that $\Gamma=\omega_0=0$, the conductivities are not affected at all by phase relaxation. In particular, the dc conductivities are divergent due to translation invariance.
\subsection{Viscosities}\label{ssec_largevisc}

Viscosities characterize momentum transport. It is sensible to talk about viscosities at timescales over which the Goldstone modes have decayed but momentum is still effectively conserved. In this regime one can simply put the momentum relaxation rate $\Gamma = 0$.

The frequency-dependent viscosities are given by the Kubo relations $\zeta(\omega) + \eta(\omega) = \lim_{k\to 0} i\omega/k^2 \times G^R_{\pi_\parallel\pi_\parallel}(\omega,k)$ and $\eta(\omega) = \lim_{k\to 0} i\omega/k^2 \times G^R_{\pi_\perp\pi_\perp}(\omega,k)$. Therefore, again using the Kadanoff-Martin method summarized in the previous section, we find
\begin{subequations}\label{eq:freq}
\begin{empheq}[left={\rm WC}:\qquad\empheqlbrace\quad, right=\qquad]{align}
\zeta(\omega) + \eta(\omega)  & 
	= \frac{ih}{\omega} + \frac{\mu+\kappa}{\Omega_\parallel-i\omega} + \zeta + \eta \, ,\\ 
\eta(\omega)   & 
	= \frac{\mu}{\Omega_\bot - i\omega} + \eta \, .
\end{empheq}
\end{subequations}
Here the inverse `internal compressibility' $h = {\rho \choose s}^i (\chi^{-1})_{ij} {\rho \choose s}^j$, where $\chi$ is the matrix of thermoelectric susceptibilities in (\ref{eq:sus}).
These then lead to the viscosities in the fluctuating WC phase
\bea\label{WC_Kubo_relax}
\label{WC_Kubo_relax_1}
\zeta^\text{WC} + \eta^\text{WC} & = &  \lim_{\omega\to 0} \text{Re} \left( \zeta(\omega) + \eta(\omega) \right) 	= \frac{\mu+\kappa}{\Omega_\parallel} + \zeta + \eta\, ,\\ \label{WC_Kubo_relax_2}
\eta^\text{WC} & = & \lim_{\omega\to 0} \text{Re} \, \eta(\omega) 
	= \frac{\mu}{\Omega_\bot} + \eta \, .
\eea
The shear and bulk viscosities are now finite but large, $\mu/\Omega_\perp$ and $(\mu+\kappa)/\Omega_\parallel$ respectively. For a CDW, the Kubo formula for the various viscosities can be written
\bea
\eta_{ijkl}^{\rm CDW} 
    = \lim_{\omega\to 0}\lim_{k\to 0}\frac{1}{\omega} \text{Im}\, G^R_{T_{ij} T_{kl}} (\omega,k)
    = \frac{\kappa_n}{\Omega} n^in^jn^kn^l + \eta^{ijkl} \,.
\eea
Recall that $\hat n$ is a unit vector in the direction of the CDW. The reduced symmetry requires the use of the Green's function of the stress tensor rather than momentum in this case.

The proliferation of dislocations has melted the lattice. In the limit that the phase relaxation rates vanish, the viscosities diverge, due to the rigidity of the translational order. The emergence of large viscosities from proliferating dislocations has been understood for some time \cite{PhysRevB.19.2457, PhysRevB.22.2514}. However, in the following \S\ref{sec_memmat} we proceed to obtain precise microscopic formulae for the $\Omega$'s in terms of the state in the dislocation cores. That result will underpin our understanding of vortex melting in \S\ref{sec:melting}.

\section{Phase relaxation from dislocation flow}\label{sec_memmat}


So far, all of our results have followed from universal hydrodynamic equations. In particular, the relaxation parameters
$\Gamma,\omega_o,\Omega_\parallel,\Omega_\bot$ are phenomenological coefficients. However, when relaxation is slow --- which is when hydrodynamics is valid --- it is possible to express these quantities in terms of microscopic correlation functions. This is achieved using the memory matrix -- closely following \cite{Davison:2016hno}, as we now describe.

\subsection{Phase relaxation with translation invariance}\label{ssec_memmat_P}

In a translation invariant system, conservation of momentum
\be
P_i = \int d^2 x \, \pi_i \,,
\ee
leads to a divergent conductivity. This divergence is removed when translations are broken, as seen in \eqref{relaxed_cond}. The memory matrix derivation of a formula for momentum relaxation $\Gamma$ has been reviewed in detail in \cite{Hartnoll:2016apf}. We shall not discuss a microscopic perspective on $\Gamma$ here, as it is not affected by spontaneous breaking of translations.

In a WC (or a CDW, see below) where translations are spontaneously broken, additional conserved operators appear: the winding of the phases across the sample
\begin{equation}\label{winding}
W_{ij} \equiv \int d^2 x \, \d_i \phi_j\, .
\end{equation}
Expectation values of the winding measure the deformation of the WC and hence conservation of these operators leads to divergent viscosities. If conservation of these operators is weakly broken, so that $\dot W_{ij} = i[H, W_{ij}]\neq 0$, then the viscosities are rendered finite by the decay rates $\Omega_\parallel$ and $\Omega_\bot$, see \eqref{WC_Kubo_relax}. These phase relaxation rates can now be obtained using the memory matrix method. The arguments below follow \cite{Davison:2016hno} closely, and are only outlined here. In particular, because conservation of the operators in \eqref{winding} is a topological effect (the winding of a phase), the operators can only be relaxed by a topological defect: dislocations. The topology of dislocations is described in Appendix \ref{app:dislocation}.

To compute $\Omega_\parallel$ and $\Omega_\bot$ it is sufficient to consider the memory matrix for the $W_{ij}$ operators in \eqref{winding}. For the computation of leading order relaxation, the memory matrix can be taken to be \cite{Davison:2016hno}
\begin{equation}\label{memmat}
M_{ij,kl} \equiv \lim_{\omega\to 0} \frac{1}{\omega}{\rm Im}\, G^R_{\dot W_{ij}\dot W_{kl}}(\omega)\, .
\end{equation}
Note the time derivatives on $W$.
This quantity can be decomposed into irreducible tensors 
\begin{equation}\label{M_irred}
M_{ij,kl} = 
	\sum_{s=0}^2 M_{(s)}P^{(s)}_{ij,kl} \, ,
\end{equation}
where the angular momentum projectors are given by 
\begin{equation}\label{projectors}
P^{(0)}_{ij,kl} = \frac{1}{2}\delta_{ij} \delta_{kl}\, , \quad
P^{(1)}_{ij,kl} = \delta_{i[k} \delta_{l]j}\, , \quad
P^{(2)}_{ij,kl} = \delta_{i(k} \delta_{l)j} - P^{(0)}_{ij,kl} \, .
\end{equation}
To leading order in slow relaxation, the memory matrix formalism then gives the relaxation rates in \eqref{WC_joseph_relax} as
\begin{equation}\label{Omega_memmat}
\Omega_\parallel 
	\simeq \frac{M^{(0)}}{\chi_{\lambda_\parallel\lambda_\parallel}}
	= (\mu+\kappa){M^{(0)}}\, , \qquad\qquad
\Omega_\bot 
	\simeq \frac{M^{(1)}}{\chi_{\lambda_\bot\lambda_\bot}}
	= \mu {M^{(1)}}\, .
\end{equation}
The remaining component $M^{(2)}$ is not independent of $M^{(0)}$ and $M^{(1)}$ and does not carry any new information. One can indeed show from \eqref{memmat} that $M^{(2)}=M^{(0)}+M^{(1)}$.

As is easily seen \cite{Davison:2016hno}, any Hamiltonian that is a local functional of $\pi$ and other operators will commute with \eqref{winding}, so that the winding is conserved: $\dot W$ and hence the $\Omega$'s are zero.
This is because the commutator ends up being a total derivative of a single-valued operator.
However, dislocation cores are in the normal phase, where the Goldstone field is undefined. In the presence of dislocations, the winding operator should be defined as
\begin{equation}
W_{ij} = \int_{\mathbb R^2 \backslash \rm cores}\!\!\! d^2 x \, \d_i \phi_j\, ,
\end{equation}
so that the integral only runs over the area where the system is in the symmetry broken phase. Any term containing $\pi_j$ in the Hamiltonian will now contribute to $\dot W_{ij}$. This follows from the nonzero commutator \eqref{WC_com} between the momentum and the phase as well as the presence of boundaries in the domain of integration. A term that is universally expected to be present in the effective low energy Hamiltonian is
\begin{equation}\label{delta_H}
\Delta H = \frac{\chi_{\pi\pi}^{-1}}{2} \int d^2 x \, \pi^2 \, .
\end{equation}
This interaction leads to the relaxation
\begin{equation}\label{fluxflowrelax}
\dot W_{ij} = - \chi_{\pi\pi}^{-1} \int_{\rm cores} d^2 x \, \d_i \pi_j\, .
\end{equation}
Terms arising in \eqref{fluxflowrelax} due to the second term in the commutation relation \eqref{WC_com} have been neglected. These are subleading both in derivatives and because they involve a higher number of operators that each have Gaussian correlations in the hydrodynamic limit.

Using the relaxation \eqref{fluxflowrelax}, equations \eqref{memmat} and \eqref{Omega_memmat} lead to a formula for the phase relaxation rate in the presence of mobile dislocations. {Assuming rotational symmetry} and the dislocations to be approximately circular for simplicity, and neglecting $\pi_i(x) \pi_j(y)$ correlators when $x$ and $y$ are in different dislocation cores, one gets
\begin{equation}\label{Omega_ff_temp}
M_{ij,kl} = \frac{n_d}{\chi_{\pi\pi}^2} \int \frac{d^2k}{(2\pi)^2}\int_{\rm core} \! d^2x\int_{\rm core} \! d^2y\, e^{ik\cdot (x-y)} \ k_i k_k \left[\lim_{\omega\to 0} \frac{{\rm Im} \, G^R_{\pi_j\pi_l}(k,\omega)}{\omega}\right]\, ,
\end{equation}
where $n_d$ is the density of free dislocations (counted without signs). Evaluation of the quantity in square brackets requires knowledge of the normal state. This quantity is a momentum density correlation function. If the dislocation cores are sufficiently large, the Green's function can be computed using normal state hydrodynamics inside the cores. Depending on the phase structure of the material, the dislocation core could either be described by a fully symmetry phase or have orientational quasi-long range hexatic order \cite{PhysRevB.19.2457,PhysRevB.22.2514}. We can easily consider both cases.

Using normal state hydrodynamics in the core, one finds
\begin{equation}\label{mm_integrand}
\lim_{\omega\to 0} \frac{{\rm Im}\, G^R_{\pi_i\pi_j}(\omega,k)}{\omega} = \frac{\chi_{\pi\pi}^2}{\eta_{\rm n}^{\rm eff}}\left[\epsilon_{ii'}\epsilon_{jj'} + \delta_{ii'}\delta_{jj'} (\ell k)^2 + \mathcal{O}(\ell k)^4\right]\frac{k_{i'}k_{j'}}{k^4}\, ,
\end{equation}
where $\eta_{\rm n}^{\rm eff}$ is an effective shear viscosity of the normal state. If the normal state were described by isotropic hydrodynamics, $\eta_{\rm n}^{\rm eff} = \eta_{\rm n}$ would simply be the shear viscosity of the normal state. For a hexatic normal phase, $\eta_{\rm n}^{\rm eff} = \eta_{\rm n} - \frac{1}{\gamma}$ where $\gamma$ is a dissipative coefficient in the hexatic hydrodynamics which characterizes diffusion of the hexatic phase\footnote{More specifically, it appears in the Josephson relation for the Goldstone field $\Theta$ of rotations (in the hexatic phase) as
\begin{equation*}
\dot \Theta = \nabla \times v + \gamma \nabla s_{\nabla \Theta} + \ldots\, ,
\end{equation*}
where $s_{\nabla\Theta} = \d f/\d(\nabla \Theta)$ is the thermodynamic source for $\nabla \Theta$.
} \cite{chaikin2000principles,PhysRevB.22.2514}. The length scale appearing in \eqref{mm_integrand} is 
\begin{equation}\label{ell}
\ell^2 \sim (\zeta+\eta)_{\rm n}^{\rm eff} \cdot \frac{\sigma_0 (\bar \kappa_0/T)-\alpha_0^2}{n^2 (\bar\kappa_0/T) - 2sn \alpha_0 + s^2 \sigma_0^2}\, .
\end{equation}
Notice that this vanishes in the Galilean limit where $\sigma_0,\, \alpha_0 \to 0$. In general, the length scale $\ell$ is intrinsic to the hydrodynamics of the normal state and should therefore be of order the mean free path (the short distance cutoff)
\begin{equation}
\ell \lesssim \ell_{\rm mfp}\, .
\end{equation}

The irreducible components \eqref{M_irred} of $M$ can be obtained by contracting \eqref{Omega_ff_temp} with the projectors \eqref{projectors}. Cutting off the momentum integral at $k_{\rm max} = 1/\ell_{\rm mfp}$, one finds that the relaxation rates \eqref{Omega_memmat} are given by
\begin{equation}\label{eq.Omega_nv}
\Omega_\parallel \sim 
	n_d \, (\mu+\kappa) \frac{r_d \ell^2}{\eta^{\rm eff}_{\rm n}\ell_{\rm mfp}}\, , \qquad\qquad
\Omega_\bot \simeq 
n_d \, \mu \frac{\pi r_d^2}{2\eta^{\rm eff}_{\rm n}}\, .
\end{equation}
Here $r_d$ is the radius of the core.
Note that the longitudinal relaxation rate $\Omega_\parallel$, which controls the relaxation of crystal compressions $\lambda_\parallel = \nabla\cdot \phi$ caused by the `climb' motion of dislocations (see Appendices \ref{app:dislocation} and \ref{ssec_modes}), vanishes in the Galilean limit. This recovers the well-known result that climb is forbidden in Galilean systems in the absence of crystal impurities \cite{PhysRevB.19.2457,PhysRevB.22.2514,Beekman:2016szb}. In systems without Galilean invariance longitudinal relaxation is allowed, but suppressed for large vortices by a factor of $\sim \ell_{\rm mfp}/r_d$ with respect to shear relaxation $\Omega_\perp$.

It was shown in the hydrodynamic treatment of the WC above that a small relaxation rate $\Omega_\bot$ leads to an anomalously large shear viscosity, see \eqref{WC_Kubo_relax_2}: 
\begin{equation}\label{large_viscosity}
\eta^{\rm WC} \simeq \frac{\mu}{\Omega_\bot} +\eta = \frac{2}{x_d} \eta_{\rm n}^{\rm eff} + \eta\, ,
\end{equation}
where $x_d=\pi r_d^2 n_d$ is the fraction of the area occupied by dislocation cores. The fact that there is a large contribution to the shear viscosity proportional to $1/x_d$ in the presence of mobile dislocations has been known for a long time \cite{PhysRevB.19.2457}; equation \eqref{large_viscosity} relates this term to a transport coefficient in the hexatic (or symmetric) state. 

The longitudinal relaxation similarly leads to large viscosities. Using \eqref{WC_Kubo_relax_1}, we have
\begin{equation}\label{large_viscosity2}
(\zeta+\eta)^{\rm WC} \simeq \frac{\mu + \kappa}{\Omega_\parallel} +\zeta + \eta\sim \eta_{\rm n}^{\rm eff}\frac{\ell_{\rm mfp}r_d}{x_d \ell^2} \sim \frac{1}{x_d} \eta_{\rm n}^{\rm eff} \cdot \frac{r_d}{\ell_{\rm mfp}} \, ,
\end{equation}
where terms subleading in the limit of small relaxation have been dropped. Although the exact expressions for $\Omega_\parallel$ and $\zeta^{\rm WC}$ depend on the microscopics through $\ell_{\rm mfp}$, \eqref{large_viscosity2} shows that if there are no other effects contributing to longitudinal phase relaxation, the anomalous bulk viscosity will generically be much larger than the anomalous shear viscosity in the fluctuating state
\begin{equation}\label{zeta_eff}
\zeta^{\rm WC} \sim \eta^{\rm WC} \cdot \frac{r_d}{\ell_{\rm mfp}} \gg \eta^{\rm WC}\, .
\end{equation}

Finally, having now obtained $\Omega$ we can turn to the `pinning frequency' $\omega_o$. This frequency is not a relaxation rate but rather appears directly as a $k=0$ susceptibility for the Goldstone mode in \eqref{f_relax}. Recall that $\omega_o$ is given in terms of
$k_o$ by \eqref{eq:pin}. The memory matrix does not offer any advantage in the computation of this quantity.
The usual field theoretic computations or energetic estimates can be applied \cite{LEE1974703, RevModPhys.60.1129, PhysRevLett.35.1399}. Appendix \ref{sec:app2} gives a fully memory matrix derivation of the optical conductivty \eqref{WC_cond}, giving a microscopic definition of all of the parameters that appear, including $\omega_o$.

\subsubsection*{Stripes}

For a CDW, a very similar story holds: the operator protecting zero frequency poles in the viscosity is
\begin{equation}
W_{nn} = \int d^2 x\, \d_n \phi_n\, .
\end{equation}
With $\Delta H$ given by \eqref{delta_H}, the memory matrix component again follows from \eqref{Omega_ff_temp}:
\begin{equation}
M_{nn,nn} = n^in^jn^kn^lM_{ij,kl} = \frac{n_d \pi r_d^2}{8 \eta_{\rm n}^{\rm eff}}\, ,
\end{equation}
where $\eta^{\rm eff}_{\rm n}$ now is an effective viscosity of the nematic (or symmetric) state inside the dislocation cores. The CDW relaxation rate is given by
\begin{equation}
\Omega = \frac{M_{nn,nn}}{\chi_{\d_n\phi_n\d_n\phi_n}} = \frac{\kappa_n}{\eta_{\rm n}^{\rm eff}} \frac{x_d}{8}\, .
\end{equation}	
%

\subsection{Phase relaxation without translation invariance}\label{ssec_memmat_noP}
When translations are broken it is important to carefully define what is meant by phase relaxation. Large viscosities are seen by probes at short enough wavelengths with
\begin{equation}\label{visc_regime}
k \gg k_o\, .
\end{equation}
At these wavelengths momentum is effectively conserved. When studying ac conductivities, one is instead interested in a regime
\begin{equation}\label{cond_regime}
k \ll k_o\, .
\end{equation}
We show below that in the presence of a pseudo-Goldstone mass, the phase relaxation rates aquire a scale dependence $\Omega\to \Omega(k)$, and take qualitatively different forms in the different regimes \eqref{visc_regime} and \eqref{cond_regime}. In particular we will see that in the limit \eqref{cond_regime} that pertains to conductivities, both relaxation rates are equal
\begin{equation}
\Omega_\parallel(k\ll k_o)
	= \Omega_\bot(k\ll k_o)\, .
\end{equation}
This was anticipated in \S \ref{sec:cond} by imposing that the longitudinal and transverse conductivities be equal for local theories. In the opposite regime \eqref{visc_regime} however, the dynamics is effectively translation invariant and the analysis of the previous section holds. In particular, one has generically 
\begin{equation}
\Omega_\parallel(k\gg k_o)
	\neq \Omega_\bot(k\gg k_o)\, ,
\end{equation}
(in fact, as discussed in the previous section, one expects $\Omega_\parallel\ll \Omega_\bot$ for Galilean invariant theories, where the `climb' motion of dislocations is suppressed) so that the large contributions to the shear and bulk viscosities in \eqref{WC_Kubo_relax} can be different.

We now prove the statements made above. When translations are broken, the free energy \eqref{f_relax} implies that the susceptibilities are modified, and the relaxation rates in \eqref{Omega_memmat} become\footnote{In the case of relaxation due to dislocations, the free energy is really integrated outside of the dislocation cores, so the susceptiblities are complicated functions of $k$. However, it is still true that in both limits \eqref{visc_regime}, \eqref{cond_regime} they take the following simple form (focusing on the shear sector for simplicity)
\begin{equation}
\chi_{\lambda_\bot\lambda_\bot}(k\gg k_o ) = 1/\mu\, , \qquad\qquad
\chi_{\phi_\bot\phi_\bot}(k\ll k_o) = 1/(\mu k_o^2)\, ,
\end{equation}
and the relaxation rates in both limits are given by
\begin{equation}
\Omega_\perp(k\gg k_o) = \chi_{\lambda_\bot\lambda_\bot} M^{(0)}\, , \qquad\qquad
\Omega_\perp(k\ll k_o) = \chi_{\phi_\bot\phi_\bot} M_{\phi_\bot\phi_\bot}\, ,
\end{equation}
with similar expressions for the longitudinal relaxation $\Omega_\parallel$, so that in either limits one recovers again \eqref{Omega_memmat} or \eqref{Omega_cond_regime}. However the crossover for $k\sim k_o$ in this case will not be described by \eqref{Omega_crossover}.
}
\begin{equation}\label{Omega_crossover}
\Omega_\parallel(k) = \left[\mu \left(1+\frac{k_o^2}{k^2}\right) + \kappa\right] M^{(0)}(k)\, , \qquad
\Omega_\perp(k) = \left[\mu \left(1+\frac{k_o^2}{k^2}\right)\right] M^{(1)}(k)\, .
\end{equation}
In the limit \eqref{visc_regime} that is relevant for viscosities, the terms containing $k_o$ vanish and one recovers \eqref{Omega_memmat} -- the analysis of \S \ref{ssec_memmat_P} then applies. In the opposite limit however one has
\begin{subequations}\label{Omega_cond_regime}
\begin{align}
\Omega_\parallel(k\ll k_o)
	&= \mu \frac{k_o^2}{k^2} M^{(0)}(k) = \frac{1}{2}\mu k_o^2 M_{\phi_\parallel\phi_\parallel}(k)\, ,  \\
\Omega_\bot(k\ll k_o)
	&= \mu \frac{k_o^2}{k^2} M^{(1)}(k) = \frac{1}{2}\mu k_o^2 M_{\phi_\bot\phi_\bot}(k)\, ,
\end{align}
\end{subequations}
where we defined 
\begin{equation}
M_{\phi_i\phi_j} (k) =  \frac{k_ik_j}{k^2} M_{\phi_\parallel\phi_\parallel}(k)  + \left[\delta_{ij} - \frac{k_ik_j}{k^2}\right] M_{\phi_\bot\phi_\bot}(k) \, .
\end{equation}
Now for any local theory the $k\to 0$ limit of $M_{\phi_i\phi_j}(k)$ will be smooth, which implies that both relaxation rates are equal in this limit.

As an application of these equations, we compute phase relaxation in the gapped theory due to dislocations. The same manipulations as in \S \ref{ssec_memmat_P} lead to 
\begin{equation}
M_{\phi_i\phi_j} = \frac{n_d}{\chi_{\pi\pi}^2} \int \frac{d^2k}{(2\pi)^2}\int_{\rm core} \! d^2x\int_{\rm core} \! d^2y\, e^{ik\cdot (x-y)} \  \left[\lim_{\omega\to 0} \frac{{\rm Im} \, G^R_{\pi_i\pi_j}(k,\omega)}{\omega}\right]\, .
\end{equation}
Taking the dislocation cores to be described by hydrodynamics with momentum relaxation (and without translational order), the leading contribution to the integrand for large cores comes from transverse momentum
\begin{equation}
\lim_{\omega\to 0} \frac{{\rm Im} \, G^R_{\pi_\perp\pi_\perp}(k,\omega)}{\omega} = \frac{\chi_{\pi\pi}^2}{\Gamma_{\rm n} \chi_{\pi\pi} + \eta_{\rm n} k^2} + \ldots \, .
\end{equation}
If momentum in the dislocations {\em diffuses} on the scale of a dislocation core $\Gamma_{\rm n} \chi_{\pi\pi} \ll \eta_{\rm n} k_v^2$, where $k_v \sim 1/r_d$, the relaxation rates are given by
\begin{equation}\label{eq:ff}
\Omega_\parallel = \Omega_\perp \sim n_d \, \mu \frac{\pi r_d^2}{4 \eta_{\rm n}} \frac{k_o^2}{k_v^2}\, ,
\end{equation}
If instead momentum is {\em relaxed} in dislocation cores $\Gamma_{\rm n} \chi_{\pi\pi} \gg \eta_{\rm n} k_v^2$, the phase relaxation rates are
\begin{equation}\label{eq:ss}
\Omega_\parallel = \Omega_\perp \simeq n_d \, \mu k_o^2 \frac{\pi r_d^2}{4 \chi_{\pi\pi}\Gamma_{\rm n}} \, .
\end{equation}
The first case (\ref{eq:ff}) is qualitatively similar to the result (\ref{eq.Omega_nv}) for dislocations in a clean theory,
with phase relaxation determined by the normal state viscosity. In the second case (\ref{eq:ss}) it is the momentum
relaxation rate of the normal state that controls phase relaxation.

\section{Applications}

\subsection{Melting of flux lattices}
\label{sec:melting}

Flux (Abrikosov) lattices form in two-dimensional superconductors when a perpendicular magnetic field is imposed. Such lattices spontaneously break translations and are thus described by the `Wigner crystal' hydrodynamics developed in the previous sections, albeit with an additional superfluid velocity mode. The transverse, shear sector of hydrodynamics is not affected by the longitudinal superfluid velocity; therefore the connection \eqref{WC_Kubo_relax_2} between shear viscosity and phase relaxation is not modified. Above the flux lattice melting temperature $T_m$, dislocations proliferate. This leads to a large shear viscosity using \eqref{large_viscosity}
\begin{equation}\label{eta_to_eta}
\eta^{\rm WC} \simeq \frac{\mu}{\Omega_\perp} \simeq \frac{2}{x_d} \eta_{s} 
	\sim \eta_{s} \exp \left[b \left(\frac{T_m}{T-T_m}\right)^\nu\right]\, ,
\end{equation}
with $\nu=0.37...$ for a solid to hexatic transition \cite{PhysRevB.19.2457}, and where $b$ is a constant of order one. The temperature dependence is controlled here by the fraction $x_d$ of the sample occupied by dislocations, which is well-understood  \cite{PhysRevB.19.2457}.

As in the previous \S \ref{sec_memmat}, the viscosity that appears on the right hand of (\ref{eta_to_eta}) is that of the state in the core of the dislocations, into which the crystal melts. In the present case of an Abrikosov lattice this is $\eta_{s}$, the shear viscosity of the superconducting state present in the dislocation cores. This viscosity controls the diffusion of transverse momentum in the superfluid core according to the Green's function
\begin{equation}
G^R_{\pi_\bot \pi_\bot}(\omega,k) = \frac{\eta_s k^2}{i\omega - (\eta_s/\chi_{\pi\pi})k^2} \,,
\end{equation}
Technically, $\eta_s$ appears in (\ref{eta_to_eta}) from inserting this Green's function into the integral \eqref{Omega_ff_temp}. This explicit connection between properties of the core and phase relaxation is our new contribution to this discussion. As we recall in figure \ref{fig_poles} in the Appendix, transverse momentum diffuses in a superfluid, unlike in a system with spatial order.

An experimentally measured strongly temperature-dependent $\eta^\text{WC}$ can be fit to \eqref{eta_to_eta} to obtain the superconducting state viscosity $\eta_s$. A simple way of measuring a shear viscosity is by studying the flow through a channel. In the case of a flux lattice, an experimental setup to achieve this has been described in \cite{PhysRevB.42.9938}, we will not review the details here. Such resistivity measurements were performed for flux liquids in NbGe in \cite{PhysRevLett.77.159} and optimally doped Bi$_2$Sr$_2$CaCu$_2$O$_8$ (BSCCO) in\cite{1742-6596-150-5-052288} (see also \cite{PhysRevLett.75.3525,Kyoso2004506}). The superconductor shear viscosity can be extracted from these experiments and in both cases is of order
\begin{equation}\label{eta_s}
\eta_s \sim [10^{-10} - 10^{-9}]\ {\rm kg\, m^{-1}\, s^{-1}}\,.
\end{equation}
For NbGe this analysis was done in \cite{PhysRevLett.77.159}. We have carried out a similar fit for BSCCO using the data in \cite{1742-6596-150-5-052288}. 

For the case of BSCCO, we can use our now cleanly established connection to the superconducting core state to give the result \eqref{eta_s} a simple physical interpretation. The superconducting state of BSCCO has long-lived Bogoliubov quasiparticles --- we will use the experimental characterization of these quasiparticles in a related  BSCCO compound \cite{PhysRevLett.90.217002} to estimate the viscosity. The viscosity of weakly interacting quasiparticles is estimated by recalling that viscosity describes transverse momentum diffusion and hence obeys $\eta_s = m n  \, D$, where $m$ is the quasiparticle mass, $n$ the quasiparticle density and $D$ the transverse momentum diffusivity. The diffusivity is estimated as $D \sim v^2 \tau$, were $v$ is the quasiparticle velocity and $\tau$ the quasiparticle lifetime. It follows that
\begin{equation}
\eta_s \sim \epsilon \tau\, , 
\end{equation}
where $\epsilon \sim n m v^2$ is the energy density. The energy density of Bogoliubov quasiparticles in $d$ spatial dimensions, and at temperatures $T \ll \Delta$, where $\Delta$ is the superconducting gap, is of order
\begin{equation}
\epsilon \sim (\Delta^3 T)^{1/2} e^{-\Delta/T} \frac{m k_F^{d-2}}{\hbar^2}\, .
\end{equation}
For the estimate we simply used a conventional Bogoliubov quasiparticle dispersion relation from BCS theory to get the density of states, neglecting nodal points. For a two-dimensional layered system with interlayer spacing $a$, one has $\epsilon_{\rm 3d} = \epsilon_{\rm 2d}/a$. Using the values $T=60\,{\rm K}$, $\Delta \sim 20\,{\rm meV}$, $\tau \sim \hbar/(20\,{\rm meV})$ and $m \sim 10 m_e$ from \cite{PhysRevLett.90.217002}, as well as $a \sim 2\, \text{nm}$ for BSCCO one obtains
\begin{equation}
\eta_s \sim 10^{-9}\ {\rm kg\, m^{-1}\, s^{-1}}\,,
\end{equation}
in good agreement with \eqref{eta_s}. This matching confirms that the coefficient of the exponential temperature dependence in \eqref{eta_to_eta} is indeed determined by precise normal state properties as discussed around \eqref{large_viscosity} above. It explains the physical origin of the magnitude of the viscosity of the phase-disordered state.

\subsection{Metal-insulator transitions in large $r_s$ devices and cuprates}
\label{sec:MIT}

Certain materials can undergo metal-insulator transitions as a function of e.g. charge density, disorder or magnetic field. There are several classes of metal-insulator transitions. We will be interested in cases where the transition is plausibly driven due to entering a fluctuating WC or CDW phase.

From \eqref{WC_cond}, the dc conductivity for a fluctuating WC is obtained to be
\be\label{eq:dcdc}
\sigma = \sigma_0 + \frac{n^2}{\chi_{\pi\pi}} \frac{\Omega }{\Omega \Gamma + \omega_o^2} \,.
\ee
As discussed in \S\ref{ssec_memmat_noP}, $\Omega_\perp = \Omega_\parallel \equiv \Omega$ in $k=0$ conductivities.
The important point for the following discussion is that the two relaxation rates $\Omega$ and $\Gamma$ enter into this expression in quite different ways.
In the following we will mostly assume that the dominant temperature dependence of the conductivity originates from the interplay of $\Omega$ and $\Gamma$, and hence we set the overall constant shift $\sigma_0 \approx 0$. A temperature-dependent $\sigma_0$ may however be important for some of the experimentally observed `insulating' phases discussed below.

\begin{figure}[h]
\centerline{
\includegraphics[width=0.6\linewidth,angle=0]{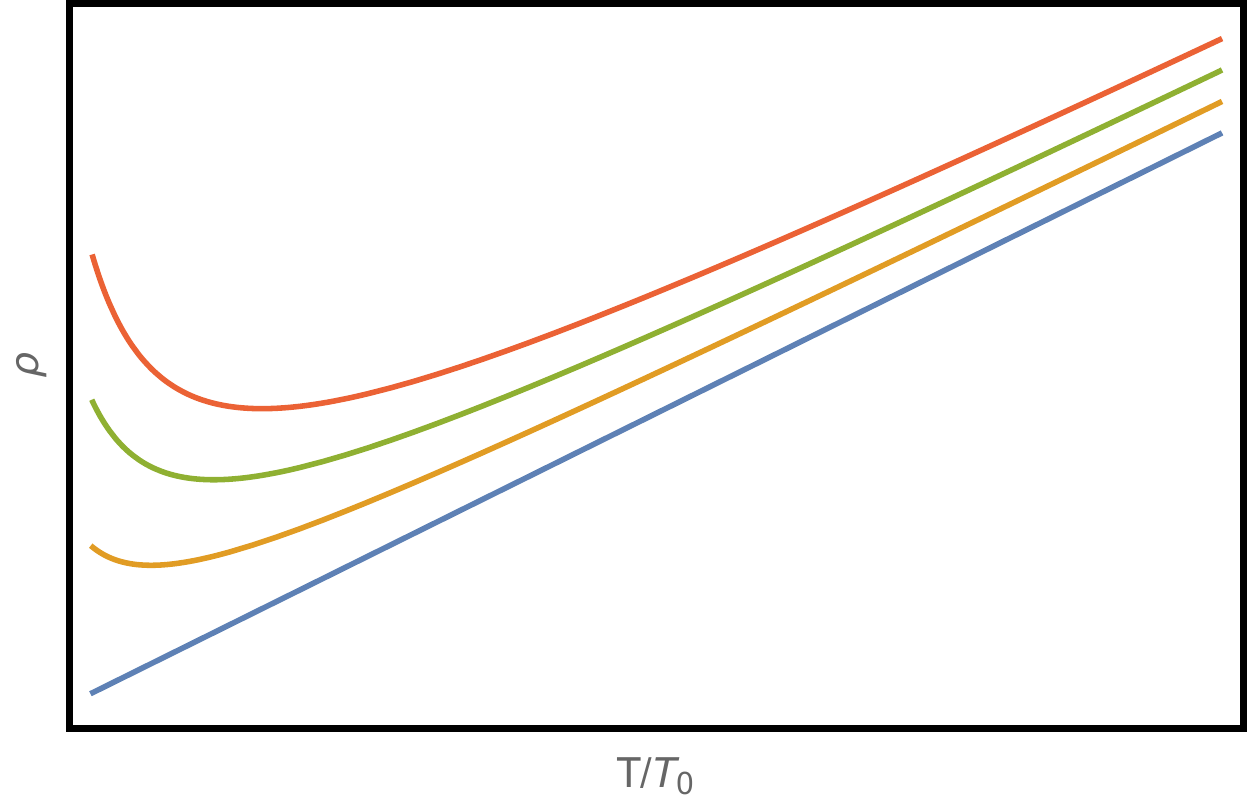}
}
\caption{\label{fig:MIT} Illustration of a metal-insulator transition. In all curves the momentum relaxation rate has been taken to be proportional to temperature $\Gamma \sim T$, while the phase relaxation rate has been modeled as $\Omega(T) \sim e^{-\sqrt{T_o/T}}$. The transition has been achieved by varying $\omega_o$ in (\ref{eq:dcdc}), with $\sigma_0 = 0$. At $\omega_o = 0$ the resistivity is metallic (lowest curve). Increasing $\omega_o$ leads to an increasingly strong resistivity upturn at low temperatures. With this choice of $\Omega(T)$ the resistivity ultimately diverges as $T \to 0$ for any $\omega_o > 0$. In practice this divergence can be cut off by a nonzero $\sigma_0$ or a low temperature saturation of $\Omega$ at a small constant value. The shape of the curves is relatively insensitive to the precise functional form of $\Omega(T)$, we have chosen this particular form for illustrative purposes only.} 
\end{figure}

At low temperatures, the relaxation rates $\Omega$ and $\Gamma$ in \eqref{eq:dcdc} will increase with temperature, while $n, \chi_{\pi\pi}$ and $\omega_o$ have only a weak temperature dependence. These assumptions can be relaxed; our immediate objective is to demonstrate the workings of a scenario for metal-insulator transitions. Far away from a WC phase, $\Omega$ will be large and temperature independent. In this limit a conventional Drude-like formula is recovered so that the resistivity $\rho = 1/\sigma \sim \Gamma$ increases with temperature and the system is metallic. Very close to an ordered WC, $\Omega$ will instead be strongly temperature dependent. Phase relaxation will be slow at low temperatures, before increasing rapidly with temperature and saturating at higher temperatures where the WC is fully melted. In the regime where $\Omega$ is increasing from a small value, the resistivity $\rho = 1/\sigma \sim 1/\Omega$ decreases with temperature and the system is insulating. The limiting case $\Omega \to 0$, with $\sigma_0 =0$, corresponds to a conventional Wigner crystal insulator. These different behaviors are illustrated in figure \ref{fig:MIT}.

A prediction of this scenario for insulating regimes where $\Omega$ is small is that there should be a peak in the optical conductivity at a nonzero frequency $\omega_o$, with width determined by $\Gamma$. This follows from the formulae in \S \ref{sec:cond} above.

\subsubsection*{Underdoped cuprates}

The temperature-dependent resistivities shown in figure \ref{fig:MIT} are reminiscent of the resistivity upturns widely observed at low temperatures in slightly underdoped cuprates. The resistivity upturn in those materials can be followed down to low temperatures upon application of a magnetic field to suppress superconductivity, e.g. \cite{PhysRevLett.77.5417, PhysRevLett.81.4720, PhysRevLett.85.638, 0295-5075-81-3-37008, Daou2009}. Without a magnetic field, the start of the upturn is visible above the superconducting temperature in some slightly underdoped samples (e.g. \cite{PhysRevLett.85.1738}) and can furthermore be extended to low temperatures by Zn substitution or irradiation, e.g. \cite{PhysRevB.39.11599, PhysRevB.51.15653,PhysRevLett.76.684}.

Undoped cuprates are Mott insulators and so of course exhibit increasing resistivity with low temperatures. The concomitant absence of 
a Drude peak and spectral weight transfer to higher energies in low doping, insulating cuprates is also well documented \cite{PhysRevB.43.7942, PhysRevB.72.054529}. However, the upturns described in the previous paragraph first appear close to optimal doping and extend into the pseudogap. This regime is now well understood to display various forms of charge density wave order (as reviewed in e.g. \cite{doi:10.1146/annurev-conmatphys-031115-011401}). The possible connection between resistivity upturns and charge density waves in cuprates has been noted in \cite{PhysRevLett.81.2132, PhysRevLett.85.1738, doi:10.1146/annurev-conmatphys-070909-104117}. Equation (\ref{eq:dcdc}) allows this scenario to be quantified in terms of observables that are in principle independently measurable.

We noted above that a characteristic signature of pinned, fluctuating charge density waves are peaks at nonzero frequency in $\sigma(\omega)$, illustrated in figure \ref{fig_conductivities}. Indeed in a Zn substituted cuprate, the low temperature Drude peak was observed to move away from $\omega = 0$ when the Zn concentration was large enough to likely lead to a resistivity upturn \cite{PhysRevLett.81.2132}. Further, optical conductivity data across the phase diagram of LSCO shows that when the start of an upturn is visible above $T_c$, significant spectral weight transfer out of the Drude peak occurs \cite{PhysRevB.43.7942}.

Similar upturns are also visible in organic and pnictide superconductors, in those cases closely tied to the presence of spin density wave (SDW) order, e.g. \cite{doi:10.1146/annurev-conmatphys-070909-104117}. Many features of our hydrodynamic analysis of incommensurate CDWs are expected to also apply to fluctuating incommensurate SDW states. SDW hydrodynamics may be a useful framework for understanding those upturns.

\subsubsection*{Large $r_s$ devices}

Metal-insulator transitions occur in devices such as Si-MOSFETs and GaAs heterostructures upon tuning the electron density. Controlling the density allows these two-dimensional systems to be tuned to large $r_s$ and hence close to WC phases \cite{RevModPhys.82.1743}. Indeed, metal-insulator transitions in these systems precisely occur as the electronic concentration is decreased (i.e. as $r_s$ is increased): the insulating regime exists in closer proximity to the WC state. Phase-fluctuating WC order may play a role over some region of the phase diagram and may be responsible for the insulating behavior. Formulae such as \eqref{eq:dcdc} allow quantitative consequences to be derived from this statement.

Previous work has suggested that the nearby WC phase is responsible for transport anomalies in these devices through the formation of a micro-emulsion of the liquid and crystalline phases \cite{KS06}. That proposal may be adiabatically connected to the one we are describing (of a phase-disordered crystal), although the experimental consequences for transport are somewhat different. In particular, we have emphasized the signatures of a fluctuating WC on optical conductivity. Such data for these devices would help corroborate the relevance of the density wave scenario.

\subsection{$\sigma(\omega,k)$}
\label{sec:sigmak}

The shift of the Drude peak to a nonzero frequency, shown in figure \ref{fig_conductivities}, is a hallmark of translational order -- it occurs whenever\footnote{In this section, we focus on the longitudinal sector of a WC and thus write $\Omega_\parallel=\Omega$ for simplicity. All of these results also apply to a CDW for transport in the direction of broken translations.}
\begin{equation}\label{eq_peakcondition}
\omega_o^2 > \frac{\Omega^3}{\Gamma + 2\Omega}\, .
\end{equation}
This phenomenon is discussed in some detail in \cite{Delacretaz:2016ivq}. Here we discuss a characteristic feature due to fluctuating density waves in 
the spatially resolved conductivity $\sigma(\omega,k)$. This is visible even if the pinning frequency $\omega_o$ is too small to produce a nonzero frequency peak at $k=0$.

\begin{figure}[h]
\centerline{
\includegraphics[width=0.4\linewidth,angle=0]{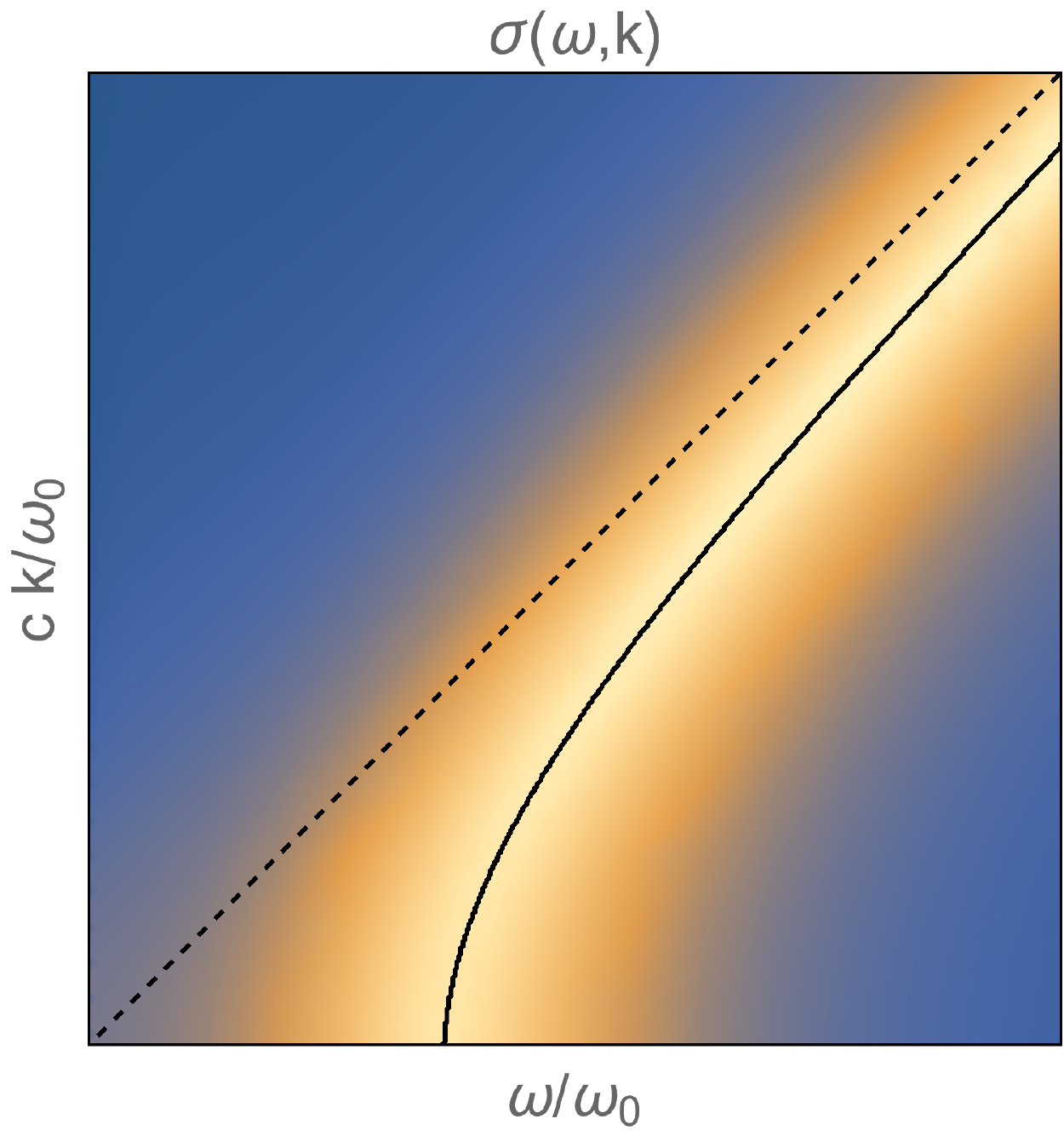}}
\caption{\label{fig_sigmawk} $\text{Re} \sigma(\omega,k)$ in a system where $\sigma(\omega,0)$ peaks at a nonzero frequency. The dashed line shows $\omega=ck$, the solid line is the position of the maximum of the conductivity. }
\end{figure}

In the scaling limit $\omega\sim k\sim \omega_o\sim \Gamma \sim \Omega$ the wavevector-dependent conductivity is
\begin{equation}\label{sigmawk}
\sigma(\omega,k)
	= \frac{n^2}{\chi_{\pi\pi}} \frac{\omega(\Omega-i\omega)}{\omega \left[\vphantom{\frac23}(\Gamma - i\omega)(\Omega-i\omega) + \omega_o^2\right] + \omega c^2 k^2 + i\Omega c_0^2 k^2} + \ldots\, ,
\end{equation}
where $c^2=(h+\mu+\kappa)/\chi_{\pi\pi}$ is the longitudinal speed of sound and $c_0^2 = h/\chi_{\pi\pi}$ is the speed of sound in the absence of a condensate (notice that setting $\Gamma = \Omega = \omega_o =0$ one recovers the sound poles $\omega=\pm ck+\ldots$). The poles of $\sigma(\omega,k)$ interpolate between those in the optical conductivity \eqref{WC_cond} at $k=0$ to the sound poles $\omega=\pm ck$ at $ck\gg \omega_o,\Omega,\Gamma$. This expression and the position of the maximum as a function of $k$ are illustrated in figure \ref{fig_sigmawk}.

\begin{figure}[h]
\centerline{
\includegraphics[width=0.95\linewidth,angle=0]{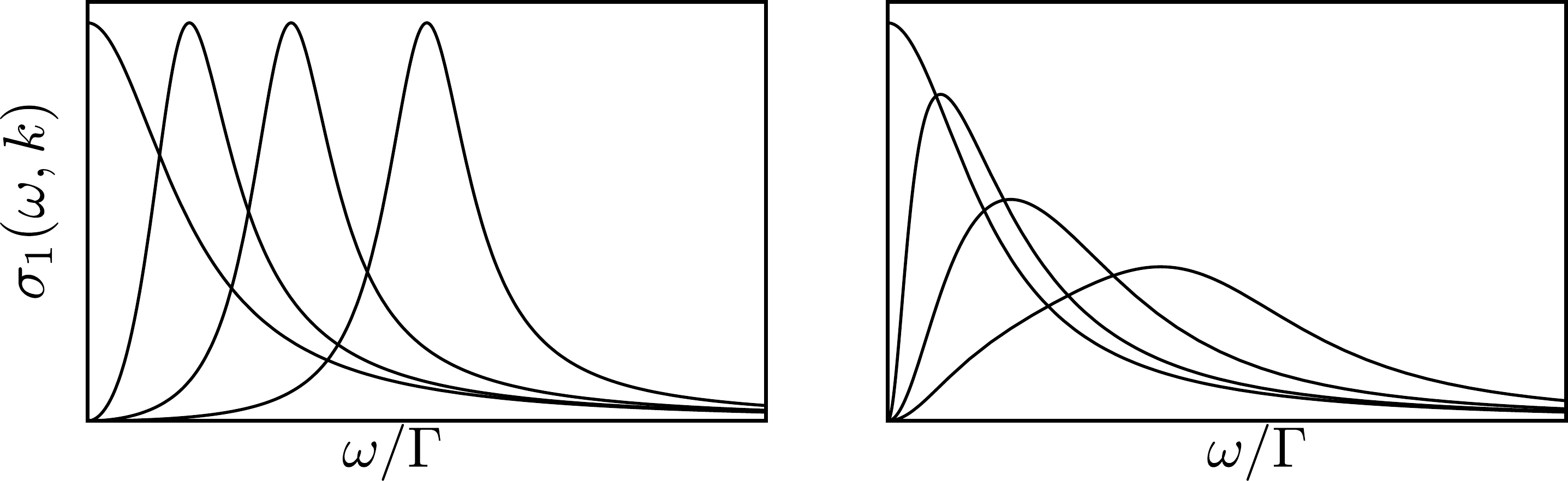}
}
\caption{\label{fig_Sigmawk2} $\text{Re} \sigma(\omega,k)$ for increasing values of $k$ for a regular metal with $\omega_o,\Omega = 0$ (left), and a system with fluctuating spatial order $\omega_o,\Omega \neq 0$ (right). In this example $\omega_o$ does not satisfy \eqref{eq_peakcondition} and is thus not large enough to cause a finite frequency peak in the optical conductivity -- as a result $\sigma(\omega,0)$ looks nearly Drude-like, but the fluctuating order is signalled by the broadening of the peak at higher wavevectors.} 
\end{figure}

Equation \eqref{sigmawk} also offers a way to detect fluctuating density wave order when $\omega_o$ is small. When \eqref{eq_peakcondition} is not satisfied, the conductivity peaks at $\omega=0$. If in addition $\omega_o^2\lesssim \Gamma\Omega$, this dc conductivity is 
\begin{equation}
\sigma_{\rm dc} \simeq \frac{n^2}{\chi_{\pi\pi}} \frac{1}{\Gamma}\, .
\end{equation}
However, the presence of fluctuating spatial order ($\Omega$) lowers and broadens the peak at nonzero wavevectors. Indeed, for wavevectors $ck\gg \Gamma,\Omega$, the conductivity peaks around $\omega\sim ck$ and the conductivity there is
\begin{equation}
\sigma^\star \simeq \sigma(c k,k) \simeq \frac{n^2}{\chi_{\pi\pi}} \frac{1}{\Gamma + \alpha \Omega}\, , 
\end{equation}
where $\alpha = (\mu + \kappa)/(h+\mu+\kappa)$. This broadening of the peak with increasing $k$ does not occur without fluctuating density waves, and is illustrated in figure \ref{fig_Sigmawk2}. At higher $k$, sound attenuation also broadens the peak.

The pole structure of $\sigma(\omega,k)$ in (\ref{sigmawk}) is also present in the retarded Green's functions for the Goldstone modes $G^R_{\lambda\lambda}(\omega,k)$ with $\lambda = \lambda_\perp$ or $\lambda_\parallel$. There are minor differences between the longitudinal and transverse channels such as different speeds of sound and phase relaxation rates. The Goldstone mode Green's functions control direct detection of translational order: they determine the singular response of the charge density structure factor $S_{nn}(\omega,Q+k)$. Here $Q$ is an ordering wavevector of the charge density waves. However, the total singular spectral weight in $S_{nn}(\omega,Q+k)$ is proportional to the density wave condensate and so is difficult to detect in fluctuating regimes. In contrast, the spectral weight of (\ref{sigmawk}) is determined by the `Drude weight' $n^2/\chi_{\pi\pi}$.


\section*{Acknowledgements}
We thank Ilya Esterlis, Andrey Gromov, Andrew Mackenzie, Akash Maharaj, and especially Steve Kivelson for helpful conversations. The work of BG is supported by the Marie Curie International Outgoing Fellowship nr 624054 within the 7th European Community Framework Programme FP7/2007-2013. LVD is supported by the Swiss National Science Foundation. AK is supported by the Knut and Alice Wallenberg Foundation. SAH is partially supported by a DOE Early Career Award.

\appendix


\section{Free energy of the Goldstone bosons}
\label{sec:hydroreview}

This appendix is a summary of known results concerning the free energy of translationally ordered phases.
When both translations are broken, there exist two Goldstone modes $\phi_i$ that satisfy
\begin{equation}\label{WC_com}
[\phi_i(x),\pi_j(y)] = i\delta^2(x-y) \left[\delta_{ij} + \d_j\phi_i\right]\, .
\end{equation}
Here $\pi_i$ is the momentum density. For a CDW, only a linear combination of these fields exists. We now build the free energy for the $\phi_i$, which will determine how these Goldstone fields appear in the hydrodynamic equations.


\subsubsection*{Crystal}

The free energy for the Goldstone modes must be invariant under all symmetries, both (nonlinearly realized) spatial and (linearly realized) residual crystalline symmetries, which is a strong constraint. We discuss the spatial symmetry first. In the WC phase there are three scalar structures, which to leading order in fields\footnote{It is possible to obtain these expressions to all order in fields using the CWZ construction \cite{Coleman:1969sm} for spatial symmetries \cite{Nicolis:2013lma,Delacretaz:2014jka}. Defining $R$ to be a rotation matrix of angle $\theta = {\rm atan}[ \nabla\times\phi/\nabla\cdot (x+\phi)]$, one finds
\begin{equation*}
U_{ij} = R_i{}^k \d_k\phi_j + R_{ij} - \delta_{ij}\, .
\end{equation*}
} are given by the components of the symmetric tensor $U_{ij}\equiv \d_{(i}\phi_{j)}$, or
\begin{equation}\label{bb}
U_{xx}=\d_x\phi_x\, , \qquad U_{yy}=\d_y\phi_y \, , \qquad\hbox{and}\qquad U_{xy}=\frac{1}{2}(\d_x\phi_y+\d_y\phi_x)\, .
\end{equation}
Each of these terms are scalars under spatial rotations, under which the fields transform as
\begin{equation}\label{rot_on_phi}
x^i\to x'^i= x^i - \varepsilon \epsilon_{ij}x^j\, , \qquad 
\phi_i \to \phi_i'(x')=\phi_i(x) + \varepsilon \epsilon_{ij}x^j\, .
\end{equation}
For a general solid, the free energy contains 6 terms at the quadratic level, built out of the scalar objects in \eqref{bb}, and can be written in the form
\begin{equation}\label{f_gen}
f=\frac{1}{2}\d_{(i}\phi_{a)}C^{ijab}\d_{(j}\phi_{b)} + \ldots\, ,
\end{equation}
with $C^{ijab}$ symmetric under $i \leftrightarrow a$, $j \leftrightarrow b$ and $(i,a) \leftrightarrow (j,b)$.

For a system with residual lattice symmetries, $C^{ijab}$ is further constrained. Extra crystalline symmetries $\subset O(2)$ act {\em linearly} on the Goldstones: $\phi^i\to R^i{}_j\phi^j$, unlike spatial rotations in \eqref{rot_on_phi}. For example, for an `isotropic solid' $C^{ijab}$ needs to be an invariant $O(2)$ tensor, which leaves only two terms in the free energy:
\begin{equation}\label{f_iso}
f= \frac{1}{2} \kappa\Tr[U]^2+ \mu \left(\Tr[U^2]-\frac{1}{2}\Tr[U]^2\right)+ \ldots\, , 
\end{equation}
where $\kappa$ and $\mu$ are called the bulk (or compression) and shear modulus respectively. This form of the free energy also applies to the triangular lattices of interest, because no other term in \eqref{f_gen} respects the 6-fold symmetry \cite{Beekman:2016szb}. The free energy can be written in Fourier space as 
\begin{equation}
f 
	= \frac{1}{2}\kappa |k\cdot \phi_k|^2 + \frac{1}{2}\mu k^2 |\phi_k|^2 + \ldots 
	= \frac{1}{2} (\kappa+\mu) |\lambda_{\parallel,k}|^2 + \frac{1}{2}\mu |\lambda_{\perp,k}|^2 + \ldots \, ,
\end{equation}
where $\lambda_\parallel = \nabla\cdot \phi$ and $\lambda_\perp = \nabla \times \phi$ parametrize the longitudinal and transverse parts of $\phi^i$ respectively. Additionally, the longitudinal field $\lambda_\parallel$ can couple to the charge and entropy densities \cite{PhysRevB.22.2514}
\begin{equation}\label{a_and_b}
f= \ldots + a \delta n \lambda_\parallel + b \delta s \lambda_\parallel\, .
\end{equation}
These couplings have little effect other than to shift certain diffusion constants and the speed of sound in the longitudinal sector, and we will take $a=b=0$ throughout.

\subsubsection*{Stripes}

In a CDW phase, only translations in the direction $\hat n$ are broken, and the single Goldstone $\phi_n$ transforms exactly like the combination $\hat n\cdot \phi$ above. The only scalar building block, to leading order in fields, is
\begin{equation}
\d_n\phi_n \equiv (\hat n\cdot \d) \phi_n\, ,
\end{equation}
which appears in the free energy as
\begin{equation}\label{f_CDW}
f = \ldots + \frac{\kappa_n}{2} \left[(\d_n\phi_n)^2 + \alpha^2 (\d_\bot \phi_n)^2\right]\, .
\end{equation}
The stiffness $\kappa_n$ of the stripe order is an elastic modulus which determines how compressible the stripes are.
In addition, symmetries allow for a coupling between $\d_n\phi_n$ and the other scalar hydrodynamic variables as in \eqref{a_and_b}. In \eqref{f_CDW} we have allowed for a term not invariant under rotations:
\begin{equation}
\d_\bot \phi_n \equiv \epsilon^{ij}n_i\d_j \phi_n\, .
\end{equation}
This term is related to an underlying anisotropy as we now explain. For most of our discussion, it is assumed that the underlying physical system is isotropic before rotations are spontaneously broken, so that the free energy must be invariant under rotations. For a CDW, this means $\alpha=0$. Physically, a small constant gradient $\d_\bot\phi_n$ is just a global rotation of the CDW, which should not cost energy in an isotropic system. The case where incommensurate CDWs do not choose their direction (because rotations are explicitly broken say by a lattice) is however of experimental interest in e.g. cuprates \cite{Gerber949}, so certain qualitative effects of taking $\alpha\neq 0$ are expected to be relevant.

\subsubsection*{No Goldstone for broken rotations}

Note that neither phase has a Goldstone $\Theta$ for broken rotations. Although it would be present in a nematic (or hexatic) phase where translations are unbroken \cite{PhysRevB.22.2514}, when at least one translation is broken it is possible to add a term to the free energy that gaps out $\Theta$:
\begin{equation}
{\rm WC:}\quad\frac{1}{2}m^2 (\Theta - \epsilon^{ij}\d_i\phi_j/2)^2 \qquad \hbox{and} \qquad {\rm CDW:}\quad\frac{1}{2}m^2 (\Theta - \d_\bot\phi_n)^2\, .
\end{equation}
These terms are invariant under rotations to leading order in fields, and at low energies pin the rotational phase $\Theta$ to the other light degrees of freedom. It is for this same reason that there are only $d$ acoustic phonons in a $d$-dimensional solid.


\section{Dislocations and their motion}
\label{app:dislocation}

\begin{figure}[h]
\centerline{
\includegraphics[width=0.45\linewidth,angle=0]{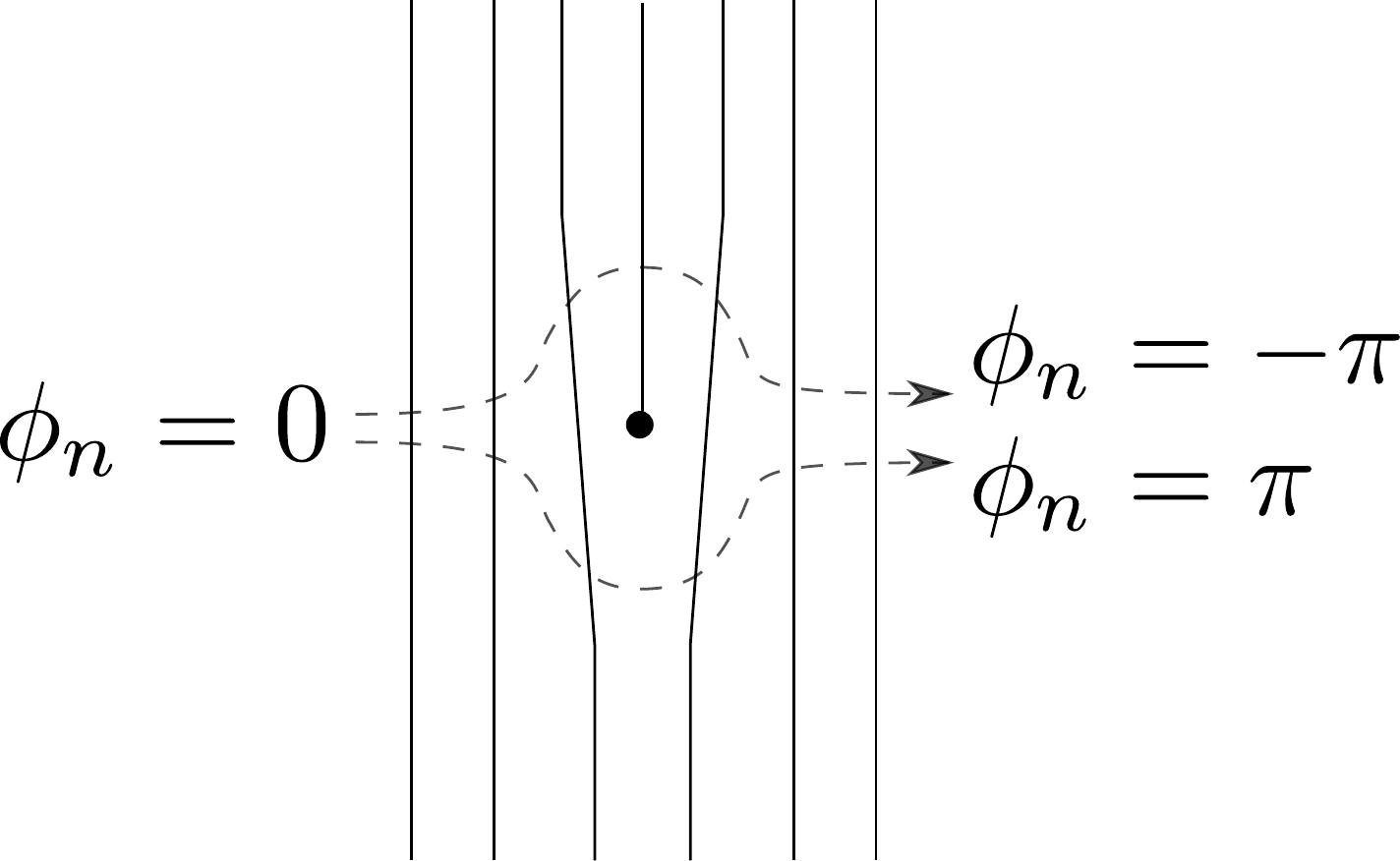}
\hspace{20pt}
\includegraphics[width=0.35\linewidth,angle=0]{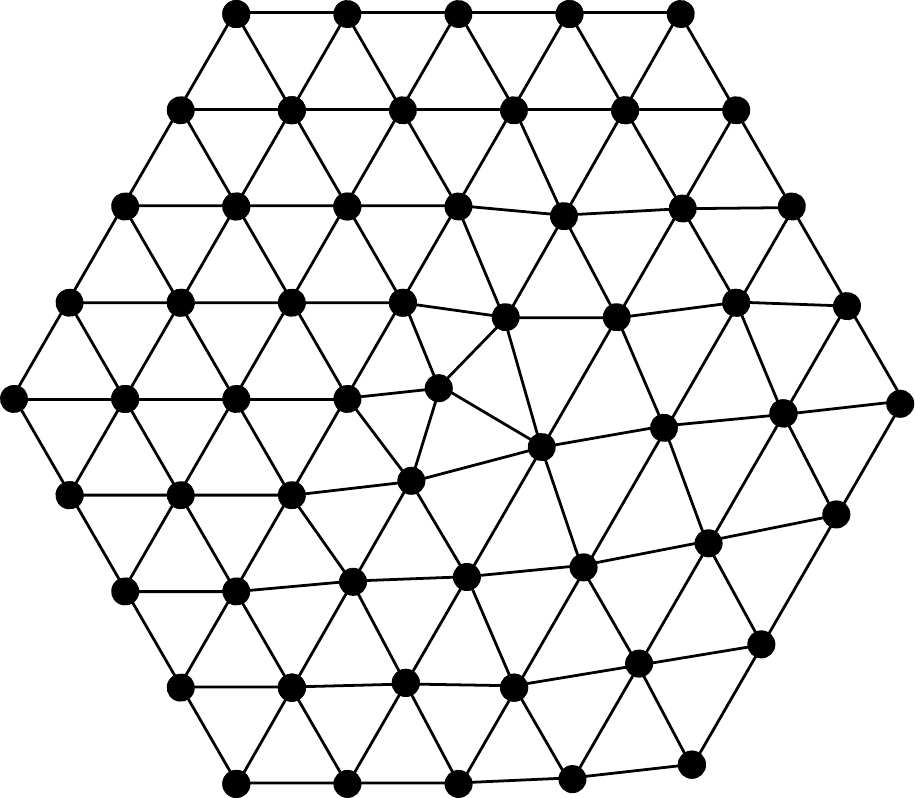}}
\caption{\label{fig_dislocation}Dislocation in a charge density wave and a Wigner crystal. The stripes in the charge density wave are lines of constant phase $Q(\hat n \cdot \vec x + \phi_n)$, where $Q\hat n$ is the CDW ordering wave vector.}
\end{figure}

Momentum relaxation ($\Gamma$) is tied to symmetry breaking, whereas phase relaxation ($\Omega$) is tied to symmetry restoration, as we now explain. Since the $\phi_i$ are phases, they can have topological defects called vortices. Vortices for the WC (or CDW) phases are dislocations, illustrated in figure \ref{fig_dislocation}. Around a simple dislocation, the phase has a winding of $2\pi$, so the gradient of the phase increases as one approaches a dislocation center, until the energy cost is too high to support such a gradient and the normal (symmetric) state is recovered.

For stripes, the presence of a gradient $\d_n \phi_n$ across the sample can only be relaxed by the transverse motion of dislocations -- this is the physical process behind the $\Omega$ term in \eqref{CDW_phase_relax}. For a WC the situation is more complicated: there are now two phases $\phi_i$ that can have any winding numbers around a dislocation. A dislocation is thus described by two integers, which specify a `Burgers vector', see e.g. \cite{chaikin2000principles, Beekman:2016szb} (for a CDW, Burgers vectors are always integer multiples of $\hat n$). These dislocations can either move parallel (`glide') or perpendicular (`climb') to their Burgers vector. The glide motion leads to a breakdown of the shear rigidity of the solid, a process captured in the shear sector of the hydrodynamics by the phase relaxation term $\Omega_\perp$ in \eqref{WC_joseph_relax_perp}. In \S \ref{ssec_largevisc} we describe how a small relaxation $\Omega_\perp$ leads to a large shear viscosity. The climb motion changes the density of the crystal lattice $\lambda_\parallel = \nabla\cdot \phi$, and thus leads to phase relaxation in the longitudinal sector \eqref{WC_joseph_relax_para}. In a Galilean system, and if all the charge is carried by the condensate, this would violate charge conservation \cite{PhysRevB.22.2514,Beekman:2016szb}. The climb motion is therefore allowed only if the diffusive mode described in \eqref{crystal_diff} is unfrozen -- it is this mode that can carry the compensating charge so that $\lambda_\parallel$ can relax without violating the continuity equation. This result will be recovered with the memory matrix computation of $\Omega_\parallel$ in \S \ref{sec_memmat}.

An appealing feature of the equations \eqref{WC_joseph_relax} describing phase relaxed WC hydrodynamics is that they require no reference to the properties of individual dislocations -- such as position or Burgers vector -- or their dynamics. The dislocations have already been `integrated out', and the most general phase relaxed Josephson relations consistent with the spatial symmetries are directly written as \eqref{WC_joseph_relax} or \eqref{CDW_phase_relax}.

\section{Hydrodynamic modes}\label{ssec_modes}

\begin{figure}[h]
\centerline{
\includegraphics[width=0.8\linewidth,angle=0]{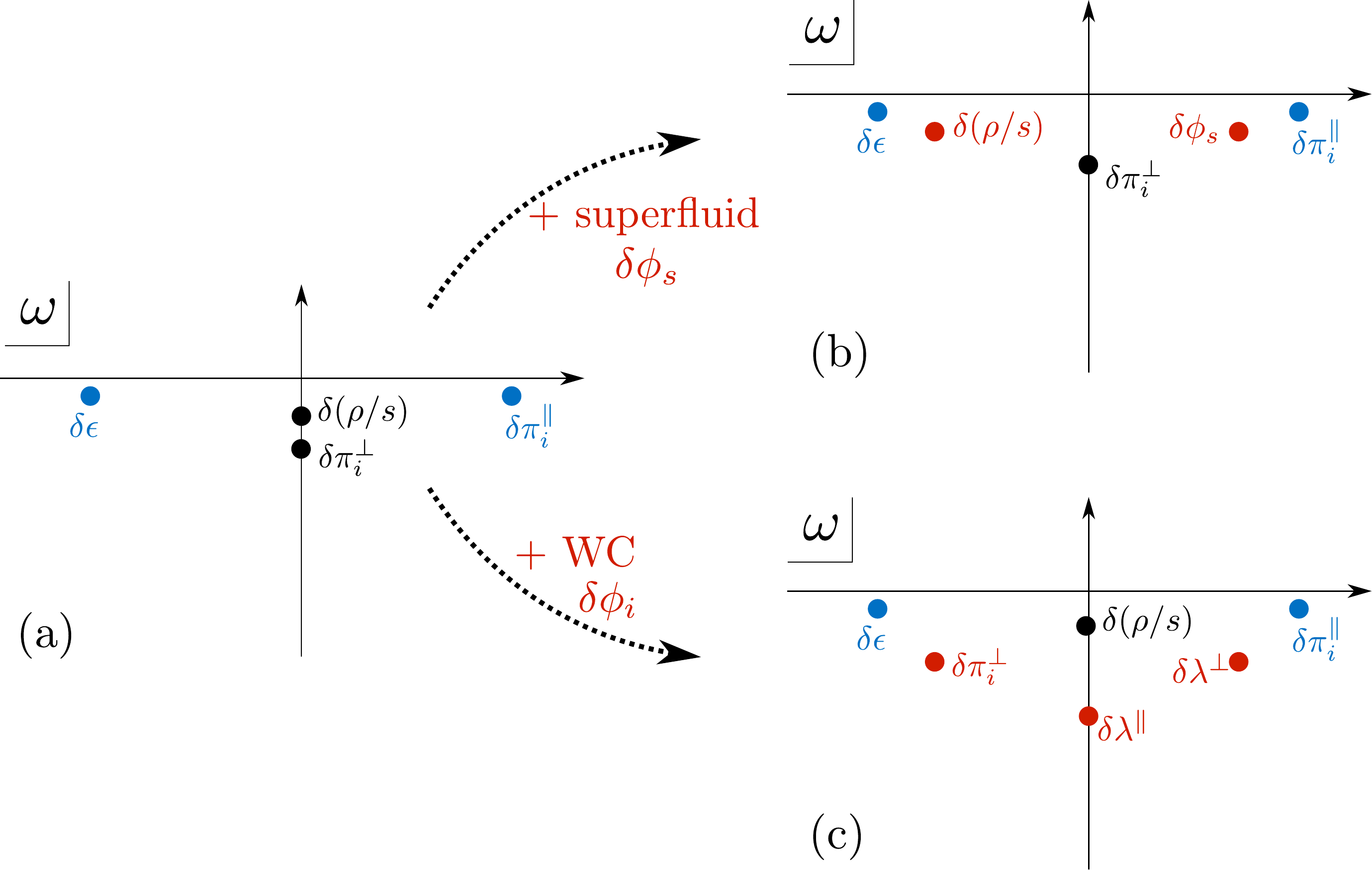}}
\caption{\label{fig_poles} Hydrodynamic modes in the complex $\omega$ plane. (a) Regular 2D hydrodynamics has longitudinal sound (in blue) carried by energy density and longitudinal momentum, and two diffusive modes carried by the transverse momentum and a combination of charge and entropy $\delta(n/s)$. (b) In the presence of a superfluid, the superfluid phase couples to the charge and lifts the corresponding diffusive mode to form second sound. (c) The transverse WC phase $\delta\lambda_\bot$ instead couples to the diffusive mode carried by transverse momentum to produce shear sound, and the longitudinal phase $\delta\lambda_\parallel$ forms its own diffusive mode. For CDW, the phase $\d_n\phi_n$ similarly couples to transverse momentum to form shear sound with a highly anisotropic speed of sound.}
\end{figure}

In the absence of momentum or phase relaxation the hydrodynamic modes solving the hydrodynamic equations of motion can be found in textbooks \cite{chaikin2000principles}. In figure \ref{fig_poles} these modes arising from translational order are compared with those arising in a superfluid.

The presence of Galilean-symmetry breaking coefficients $\sigma_0, \alpha_0, \gamma_1$ lead to small changes to the textbook formulae, but the modes are qualitatively the same. The only exception concerns the `crystal diffusion mode', with diffusion constant
\begin{equation}\label{D_CDW}
D_{\varphi}\Big|_{\sigma_0=\alpha_0=\gamma_1=0} \simeq (\kappa + \mu)\left[\xi_\parallel - \frac{ \gamma_2^2}{\bar\kappa_0/T}\right]+ \mathcal{O}(\kappa + \mu)^2\, ,
\end{equation}
where the last expression was simplified by taking the Galilean limit. In general it is carried by the hydrodynamical variable 
\begin{equation}\label{crystal_diff}
\begin{split}
\varphi 
	= & 
	\left[n (\bar\kappa_0/T) - s\alpha_0 + s(n\gamma_2 - s\gamma_1)\right]\delta n \\
	&+\left[s \sigma_0 - n\alpha_0 + n(s\gamma_1 - n\gamma_2)\right]\delta s \\
	&+\left[n^2 (\bar\kappa_0/T) - 2sn \alpha_0 + s^2 \sigma_0\right]\lambda_\parallel
	+ \mathcal{O}(\kappa + \mu) + \mathcal{O}(k)\, ,
\end{split}
\end{equation}
which has a simple diffusive Green's function
\begin{equation}
G^R_{\varphi\varphi}(\omega,k) = \frac{\chi_{\varphi\varphi}D_\varphi k^2}{i\omega - D_\varphi k^2}\, . 
\end{equation}
In the Galilean limit, and taking $\gamma_2\to 0$, the hydrodynamical variable $\varphi$ simply becomes
\begin{equation}\label{eq:tata}
\varphi \simeq \delta n + n \lambda_\parallel = \delta n + n \,(\nabla\cdot \phi)\, ,
\end{equation}
which was interpreted as a defect (or impurity) density in \cite{PhysRevB.22.2514}. More generally, it corresponds to the charge that is not transported by compressions of the crystal/charge density wave. One should read \eqref{eq:tata} as the total charge fluctuation $\delta n$ {\it minus} the charge fluctuation $- n \,(\nabla\cdot \phi)$ due to compression of the density wave. If all charge is carried by the density wave condensate, so that all the higher derivative terms in the Josephson relation \eqref{joseph} are zero, then \eqref{D_CDW} shows that the mode freezes $\omega(k) = -i D_{\varphi}k^2 = 0$. The presence of this diffusive mode in more general, non-Galilean invariant, circumstances is what allows the `climb' motion of dislocations that is discussed in \S \ref{ssec_memmat_P}.

The incorporation of slow phase relaxation into hydrodynamics has several effects. Firstly, the crystal diffusion mode described above is damped.
Secondly, the shear sound mode becomes one damped mode (for $\lambda_\perp$) and one diffusive mode (for transverse momentum $\pi_\perp$). The diffusive mode has a large diffusivity $D_\perp = \mu/(\Omega_\perp \chi_{\pi\pi})$, which is responsible for the large shear viscosity (\ref{WC_Kubo_relax_2}). Thirdly, longitudinal sounds survives but acquires a large attenuation $\Gamma_s = (\mu + \kappa)/(\Omega_\parallel \chi_{\pi\pi})$, which is responsible for the large bulk viscosity (\ref{WC_Kubo_relax_1}).

Relaxation of momentum then further damps the transverse momentum diffusive mode while the longitudinal sound mode becomes one damped mode (for $\pi_\parallel$) and one diffusive mode (for energy density $\epsilon$).

These different regimes can be accessed at different wavectors, depending on the associated timescale $\overline k^2 \equiv D k^2$, where $D$ represents any of the diffusion or sound attenuation constants in the unrelaxed theory. The regimes are illustrated in the following table:

\begin{table}[h]
\begin{center}
\begin{tabular}{|c||c|c|c|}
\hline
& \specialcell[t]{\sc  Incoherent Hydro\\ $\overline k^{\, 2} \ll \Gamma$}  & \sc \specialcell[t]{\sc  Regular Hydro\\ $\Gamma \ll \overline k^{\, 2} \ll \Omega$} & \specialcell[t]{\sc  Ordered Hydro\\ $\Omega \ll \overline k^{\, 2}$} \\
\hline\hline
\specialcell[h]{\sc  Thermoelectric \\ \sc Diffusion} & Diffusive & Diffusive & Diffusive \\
\hline
\sc Longitudinal Sound & \specialcell[h]{1 Diffusive ($\epsilon$) \\ 1 Damped ($\pi_\parallel$)} &Sound & Sound \\
\hline
\sc Shear Sound & Damped & \specialcell[h]{1 Diffusive ($\pi_\perp$) \\ 1 Damped ($\lambda_\perp$)} & Sound \\
\hline
\sc Crystal Diffusion & Damped & Damped & Diffusive \\
\hline
\end{tabular}
\end{center}
\label{tab:modes}
\end{table}%

\section{Entropy production}\label{app_S}

Positivity of entropy production in hydrodynamics leads to constraints on the dissipative coefficients appearing in the constitutive and Josephson relations. These constraints are derived in this section. For completeness we give here the leading higher derivative corrections to the Josephson relations. For the WC
\begin{subequations}\label{joseph}
\begin{align}
\dot \lambda_\parallel + \Omega_\parallel \lambda_\parallel \label{joseph_L}
	&= \nabla\cdot v + \gamma_1 \d^2 \mu_e + \gamma_2 \d^2 T + \xi_\parallel (\kappa + \mu) \d^2 \lambda_\parallel + \ldots \, ,\\
\dot \lambda_\perp +  \Omega_\perp \lambda_\perp  \label{joseph_T}
	&= \nabla \times v + \xi_\perp \mu \d^2 \lambda_\perp + \ldots\, ,
\end{align}
\end{subequations}
and for the CDW
\begin{equation}\label{joseph_CDW}
\dot \phi_n + \Omega \phi_n= v_n + \gamma_1 \d_n \mu_e + \gamma_2 \d_n T + \kappa_n\left(\xi_\parallel \d_n^2 + \xi_\perp \d_\perp^2\right) \phi_n + \ldots\, .
\end{equation}
Once translations are broken, and the pseudo-Goldstone modes are massive, there is a nonlocal relation between the source and displacement: e.g. $\pa^2 s_\perp = \mu (\pa^2 - k_o^2)\lambda_{\perp}$. Thus higher derivative terms in the source can introduce 0-derivative terms in the displacement. The phase relaxations $\Omega_{\parallel,\perp}$ are defined as the coefficients of the 0-derivative $\lambda_{\parallel,\perp}$ terms in the Josephson relation. This has been implemented in the above equations by writing derivatives of $\lambda_{\parallel,\perp}$ on the right hand side rather than derivatives of $s_{\parallel,\perp}$.

In the remainder of this appendix, for simplicity, we consider only one-dimensional charge order -- generalization to higher dimensional CDW or WC systems is straightforward. Indices will be suppressed, so that $j=j_x, \, \tau=\tau_{xx}, \, \nabla \phi = \d_x\phi_x$, etc. We will allow for both momentum ($\Gamma,\, k_o$) and phase ($\Omega$) relaxation. As usual, we study linearized hydrodynamics around a background with $\mu,\, T\neq 0$ but $v,\, \nabla\phi=0$. Up to second order in perturbations around this background, one can define a heat current as \cite{chaikin2000principles}
\begin{equation}
j^Q = j^E - \mu j + (p-\tau) v + \ldots\, ,
\end{equation}
where the pressure is given by $p = -\epsilon + \mu n + s T$.
Using the first law for a relaxed CDW 
\begin{equation}
d \epsilon = T d s + \mu d n + v d\pi + \kappa (\nabla\phi d\nabla\phi + k_o^2 \phi d \phi)\, ,
\end{equation}
the change in entropy density is, up to second order in perturbations,
\begin{align}
\dot s
	&= \frac{1}{T} \left[\dot \epsilon - \mu \dot n - v\dot \pi - \kappa (\nabla\phi \nabla\dot\phi + k_o^2 \phi \dot \phi)\right]\\ \nonumber
	&= - \nabla (j^Q/T) - \frac{1}{T}\left[(j^Q/T - s v)\nabla T + (j-n v)\nabla \mu + (\tau - p)\nabla v -\Gamma\chi_{\pi\pi} v^2- s_{\nabla \phi} \nabla\dot\phi\right]\, ,
\end{align}
where we in the last step used the (relaxed) conservation equations, and $s_{\nabla\phi} = \d \epsilon/\d(\nabla\phi)$ is the thermodynamic source for $\nabla\phi$. Using the constitutive relations \eqref{consti} and the Josephson relation \eqref{CDW_phase_relax}, one finds that positivity of entropy production
\begin{equation}
\dot s + \nabla (j^Q/T) \geq 0 \,,
\end{equation}
requires the following matrix to be positive definite for all wavevectors $k$
\begin{equation}
M(k)=
\left(\begin{array}{cccc}
\sigma_0 k^2 &\alpha_0 k^2&\gamma_1 k^2 &0 \\
\alpha_0 k^2 &\frac{\bar\kappa_0}{T} k^2&\gamma_2 k^2 &0 \\
\gamma_1 k^2 & \gamma_2 k^2 &\left(\frac{\Omega}{\kappa} + \xi k^2\right) \frac{k^2}{k^2 + k_o^2} &0 \\
0&0& 0& \Gamma \chi_{\pi\pi} + \zeta k^2 \\
\end{array}\right)\, .
\end{equation}
In addition to the constraints $\sigma_0,\,\bar\kappa_0,\,\zeta,\,\Gamma \geq 0$ and $\alpha_0^2\leq \sigma_0\bar\kappa_0/T$ that are standard in hydrodynamics, the CDW parameters must satisfy fairly non-trivial constraints, e.g.
\begin{equation}
\gamma_1^2 \leq \sigma_0 \xi\, \frac{k^2 + \frac{\Omega}{\kappa\xi}}{ k^2 + k_o^2}\, , \quad \hbox{for all }k.
\end{equation}
This leads to three different cases:
\begin{equation}\label{eq:appgamma}
\left\{
\begin{array}{ll}
\gamma_1^2 \leq \sigma_0\xi	\, ,
	& \hbox{if } k_o= 0\, ,\\
\gamma_1 = 0\, ,	
	& \hbox{if } k_o\neq 0 \hbox{ and }\Omega = 0\, ,\\
\gamma_1^2 \leq \sigma_0\min \left[\xi,\, \frac{\Omega}{\kappa k_o^2}\right]\, ,	
	& \hbox{if } k_o,\, \Omega \neq 0\, ,\\
\end{array}
\right.
\end{equation}
Similar constraints must be satisfied by $\gamma_2$, with $\sigma_0\to \bar\kappa_0/T$.

\section{Conductivities from the memory matrix}
\label{sec:app2}

The relaxed hydrodynamic conductivities \eqref{WC_cond}, \eqref{CDW_cond} can be obtained from the memory matrix. We will give an outline of the derivation, see \cite{Hartnoll:2016apf} for more details of the formalism. For simplicity, we will focus on a CDW in one dimension -- the treatment is identical for a triangular WC in two dimensions, and more complicated for a WC of reduced residual symmetry or a CDW in two dimensions. In order to obtain the electrical conductivity, it is sufficient to keep the electric current and the slow operators it overlaps with: $\{J,\,P,\, \phi\}$. The crucial difference with an ordinary metal is that in a state with translation order, the Goldstone $\phi$ for translations (sliding mode) must be included in the treatment. In this section, all operators are averaged over all space, e.g.~$J=\int dx \,j(x)$. The memory matrix formalism gives the following expression for the conductivity:
\begin{equation}\label{mm_gen}
\sigma(\omega) = \sum_{AB	}\chi_{JA}\left(\frac{1}{i\omega\chi - M(\omega) - N}\right)_{AB}\chi_{BJ} \,,
\end{equation}
where the sum is chosen to run over $A,B\in \{J,\,P,\, \phi\}$. The matrix of susceptibilities is symmetric and given by
\begin{equation}
\chi = 
\left(\begin{array}{ccc}
\chi_{jj} & \chi_{j\pi} & 0\\
\chi_{j\pi}&\chi_{\pi\pi}&0\\
0&0&\chi_{\phi\phi}
\end{array}\right)\, ,
\end{equation}
where we noted that $\chi_{j\phi}=\chi_{\pi\phi} = 0$ by time-reversal symmetry. Comparison with the hydrodynamic expressions gives $\chi_{j\pi} = n$. When translations are an exact symmetry the Goldstone susceptibility diverges at $k=0$, but the mass the Goldstone acquires when translations are broken regularizes the susceptibility%
\footnote{
When $k_o=0$ one must therefore work with the operators $J_k,\, P_k,\, \phi_k$ at nonzero $k$ -- so that the Goldstone susceptibility $\chi_{\phi_{-k}\phi_k}=1/(\kappa k^2)$ is finite. One can take $k\to 0$ at the end of the computation. 
}%
, and the free energy \eqref{f_relax_CDW} gives $\chi_{\phi\phi}=1/(\kappa k_o^2)$. The $N$ matrix is anti-symmetric, and $N_{AB}$ vanishes for operators $A,\,B$ with the same signature under time-reversal. One is left with
\begin{equation}
N = 
\left(\begin{array}{ccc}
0 & 0 & N_{j\phi}\\
0&0&1\\
-N_{j\phi}&-1&0
\end{array}\right)\, ,
\end{equation}
where we used
\begin{equation}
N_{\pi\phi} = -i \langle [P,\phi]\rangle = 1\, .
\end{equation}
Finally, all matrix elements of $M$ can be non-zero in general. We will assume momentum relaxation is weak, so that
\begin{equation}
M_{\pi\pi} \equiv \Gamma \chi_{\pi\pi}
\end{equation}
is small. It is possible to show that $M_{\pi A} = O(\Gamma)$. In the presence of phase fluctuations, we have
\begin{equation}
M_{\phi\phi} \equiv \Omega \chi_{\phi\phi} = \frac{\Omega}{\kappa k_o^2}\, .
\end{equation}
In the scaling limit $\omega \sim \Gamma \sim \omega_o \sim \Omega \sim \epsilon$ (where $\omega_o^2 = \kappa k_o^2/\chi_{\pi\pi}$), to leading order in $\epsilon$ the memory matrix formula \eqref{mm_gen} gives
\begin{equation}\label{eq:concon}
\sigma(\omega) = \frac{n^2}{\chi_{\pi\pi}} \frac{(\Omega-i\omega)}{(\Omega-i\omega)(\Gamma-i\omega)+\omega_o^2} + \ldots \, ,
\end{equation}
in agreement with \eqref{WC_cond}. The incoherent term $\sigma_o$ in (\ref{WC_cond}) is generically subleading in the scaling limit we have taken here. Such incoherent terms can be made to appear at leading order in the memory matrix computation if the incoherent spectral weight is simultaneously scaled to be large \cite{Hartnoll:2016apf}.

Finally, note that the conductivity (\ref{eq:concon}) satisfies the usual sum rule
\begin{equation}
\int_0^\infty d\omega \, {\rm Re}\, \sigma(\omega) = \frac{\pi}{2} \frac{n^2}{\chi_{\pi\pi}}\,. 
\end{equation}

\providecommand{\href}[2]{#2}\begingroup\raggedright\endgroup

\end{document}